\documentclass[12pt]{article}\usepackage[hyperfootnotes=false]{hyperref}
   \usepackage{epsfig}
      \usepackage{amsmath}
      \usepackage{amssymb}
  \usepackage{graphicx}
  \usepackage{color}
  \newlength{\abstractwidth}
  \renewcommand{\thefootnote}{\fnsymbol{footnote}}
  \renewcommand{\thanks}[1]{\footnote{#1}} % Use this for footnotes
  \newcommand{\starttext}{
  \setcounter{footnote}{0}
  \renewcommand{\thefootnote}{\arabic{footnote}}}
  \renewcommand{\theequation}{\thesection.\arabic{equation}}
  \newcommand{\be}{\begin{equation}}
  \newcommand{\bea}{\begin{eqnarray}}
  \newcommand{\eea}{\end{eqnarray}}
  \newcommand{\beq}{\begin{equation}}
  \newcommand{\ee}{\end{equation}}
  \newcommand{\eeq}{\end{equation}}

  \def\ba{\begin{eqnarray}}
  \def\ea{\end{eqnarray}}

  \def\12{{1 \over 2}}

  \def\ra{\rangle}
  
 \def\simleq{\; \raise0.3ex\hbox{$<$\kern-0.75em
      \raise-1.1ex\hbox{$\sim$}}\; }
 \def\simgeq{\; \raise0.3ex\hbox{$>$\kern-0.75em
      \raise-1.1ex\hbox{$\sim$}}\; }

\def\O2{\Omega_2}

\def\bi{\begin{itemize}}
  \def\ei{\end{itemize}}

\def\S{Schwarzschild}
\def\sc{\setcounter{equation}{0}}

\def\W{$\Omega$}
\def\W'{$\Omega$}

\def\V{\Omega}
\def\V'{\Omega}
\def\dof{degrees of freedom  }

\def\er{Einstein-Rosen bridge}
\def\ers{Einstein-Rosen bridges}

\def\nref#1{(\ref{#1})}

\begin{document}
  \renewcommand{\theequation}{\thesection.\arabic{equation}}

\begin{titlepage}
  \rightline{}
  \bigskip

  \bigskip\bigskip\bigskip\bigskip

    \bigskip
\centerline{\Large \bf {  Cool horizons for entangled black holes }}
\centerline{\Large \bf {
}}
    \bigskip

  \begin{center}
 \bf {{Juan Maldacena$^1$ and  Leonard Susskind$^2$}}
  \bigskip \rm
\bigskip

$^1$ Institute for Advanced Study,  Princeton, NJ 08540, USA
\\ ~~
\\
$^2$ Stanford Institute for Theoretical Physics and  Department of Physics, \\
 Stanford University,
Stanford, CA 94305-4060, USA \\
\rm

\bigskip
\bigskip

\vspace{2cm}
  \end{center}

  \bigskip\bigskip

 \bigskip\bigskip
  \begin{abstract}

General relativity contains solutions in which two distant black holes are connected through
the interior via a wormhole, or Einstein-Rosen bridge. These solutions can be interpreted as
maximally entangled states of two black holes that
form a complex EPR pair. We suggest that similar bridges might be present for more general entangled states.

In the case of entangled black holes one can formulate versions of the
AMPS(S) paradoxes and resolve them. This suggests possible resolutions of the
firewall paradoxes for more general situations.

 \medskip
  \noindent
  \end{abstract}

  \end{titlepage}

    \starttext \baselineskip=17.63pt \setcounter{footnote}{0}
  \tableofcontents

  \sc
  \section{Introduction}

Spacetime locality is one of the cornerstones in our present understanding of physics.
By locality we mean the impossibility of sending signals faster than the speed of light.
Locality {\it appears} to be challenged both by quantum mechanics and by general relativity.
Quantum mechanics gives rise to  Einstein Podolsky Rosen
(EPR) correlations \cite{Einstein:1935rr}, while general relativity allows solutions to the  equations of motion that connect far away regions through
relatively short ``wormholes'' or Einstein Rosen bridges \cite{Einstein:1935tc}.
 It has long been understood that these two effects do { \it not}
 give rise to real violations
of locality. One cannot use
EPR correlations to send information faster than the speed
of light. Similarly,
 Einstein Rosen bridges do not allow us to
 send a signal from one asymptotic
region to the other, at least when suitable positive
energy conditions are obeyed  \cite{Fuller:1962zza,Friedman:1993ty,Galloway:1999bp}.
 This is sometimes stated as saying that Lorentzian wormholes are not
 traversable\footnote{ This can be shown using the   integrated null energy condition \cite{Friedman:1993ty,Galloway:1999bp},
which is a correct condition in the classical theory.
It can be   violated by a small amount in the quantum theory,
   but, as far as we know, not by
   enough to make wormholes traversable. We will assume that wormholes remain un-traversable in
   the quantum theory. If this were not true, the ER=EPR connection would be wrong. }.

Here we will note that these two effects are actually connected. We argue that
the Einstein Rosen bridge between two black holes is created by EPR-like correlations between the
microstates of the two black holes. This is based on previous observations  in \cite{Israel:1976ur,Maldacena:2001kr}.
 We call this the ER = EPR  relation.
In other words, the ER bridge is a special kind of EPR correlation in which the EPR correlated quantum systems have a weakly coupled Einstein gravity description. It is
also special because the combined state is just one particular entangled state out of many possibilities.
We note that black hole pair creation in a magnetic field ``naturally'' produces a
pair of black holes in this state.
It is very
tempting to think that {\it any} EPR correlated system is connected by some sort of ER bridge, although in general the bridge may be a highly quantum object
 that is yet to be independently defined. Indeed, we speculate that  even the simple  singlet state of two spins is connected
by a (very quantum)  bridge of this type.

In this article we explain the reasons for expecting such a connection. We  also explore
some of the implications of this point of view for the black hole information problem, in its
AMPS(S)\cite{Almheiri:2012rt,Almheiri:2013hfa}  form. See \cite{Braunstein:2009my,Mathur:2009hf,Giddings:2012dh} for some earlier work and
\cite{Almheiri:2013hfa} for a
more complete set  of references.   See \cite{Papadodimas:2012aq} for a proposal to describe interiors
that is similar to what we are saying here\footnote{   For other work trying to connect two sided black
holes with the AMPS paradox see \cite{Avery:2013exa}.}. 
 
 The first point is that two black holes that are far away but connected by an
ER bridge provide an existence proof of a black hole that is maximally entangled with a second
distant system, but which nevertheless has a smooth horizon.
 On the other hand,  AMPS \cite{Almheiri:2012rt} suggested that the smoothness of the interior
will be destroyed once the black hole becomes entangled with another system; the second system
being the radiation in their case.
If an observer collects the radiation, then, with a powerful enough quantum computer\footnote{  We are ignoring possible limits on quantum computation \cite{Harlow:2013tf}.}, she could
collapse it into a second black hole which is perfectly entangled with the first. In addition, by operations solely on her side, she can put the
pair of black holes  in the special state that produces the smooth ER bridge.
Thus we argue that the action of a quantum computer on the radiation can produce a
state where the horizon is smooth.

 Consider the case of two very distant black holes which are entangled in the state that produces the
 ER bridge. Bob is stationed at one (the near black hole) and Alice is at the other (far black hole). Alice has a powerful quantum computer that can act on her black hole. Then it is possible for Alice to  send messages to Bob through the \er. Bob cannot receive the messages as long as he is outside his horizon, but as soon as he passes through the horizon he can receive the messages. If Alice chooses,  she can create a firewall at Bob's end. The original AMPS experiment can be restated as sending such a message.
 We see that  actions on the radiation are not innocuous; they can affect what Bob feels when he falls
 through the horizon.

 In acting with her quantum computer on the radiation, Alice has created a   very special state.  What if she does not act on the radiation at all?. A naive picture is that
 the radiation would be  connected by very quantum ER bridges to itself and also to the black hole horizon.
 Thus, whether the black hole horizon is smooth or not depends on how these quantum bridges connect to
 form the big classical geometry outside the horizon of the first black hole. If we trust the
 equivalence principle, then we would conclude that the bridge remains big and classical in the
 interior of the black hole.  However, we do not have an independent argument for its smoothness.

\sc
\section{Einstein Rosen Bridges}
  \subsection{AdS Black Holes}

\begin{figure}[h!]
\begin{center}
\includegraphics[scale=.8]{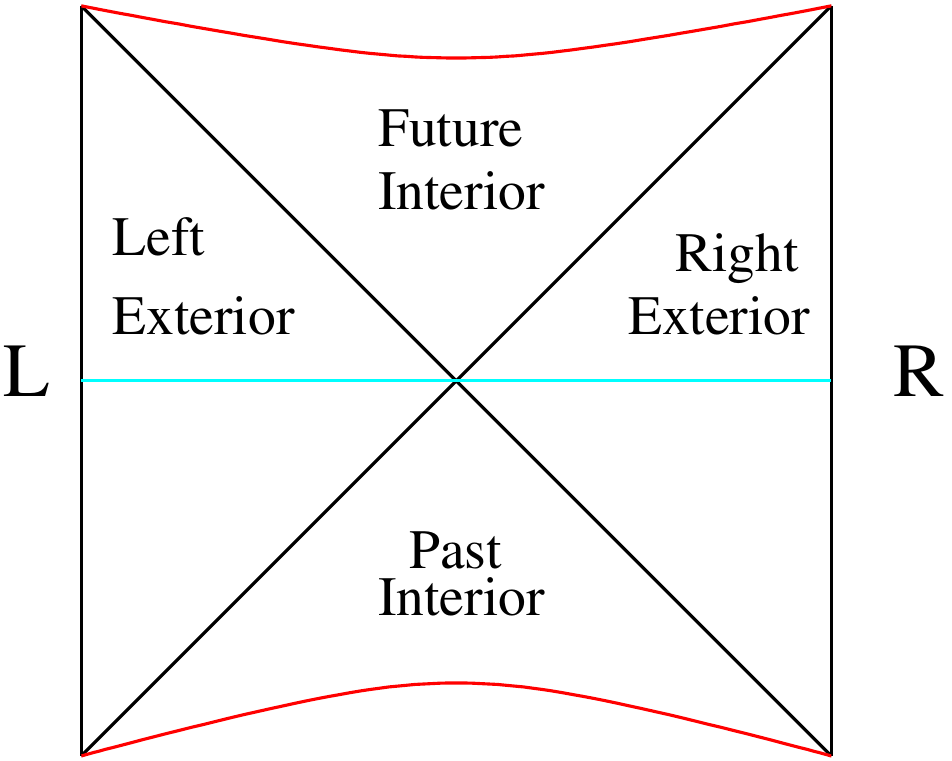}
\caption{ Penrose diagram of the eternal black hole in $AdS$. 1 and 2, or Left and Right,
 denote the two boundaries and
the two CFT's that the system is dual to.   }
\label{1}
\end{center}
\end{figure}

Einstein-Rosen bridges and their relation with entanglement is most rigorously understood in the ADS/CFT framework. Consider the eternal AdS-\S \ black hole whose  Penrose diagram is shown in figure \ref{1}.
This diagram displays the two exterior regions and two interior regions. It is important not to
confuse the future interior with the left exterior. Sometimes the left exterior is referred colloquially
as the ``interior'' of the right black hole, but we think it is important not to do that.
Note that no signal from the future
interior can travel to either  of the two exteriors.

The system is described by two identical uncoupled CFTs defined on disconnected boundary spheres. We'll call them the Left and Right sectors. The energy levels of the QFT's  $E_n$ are discrete. The corresponding eigenstates are denoted $|n\ra_L, \ |n\ra_R.$ To simplify the notation the tensor product state $|n\ra_L \otimes|m\ra_R $ will be called $|n, \ m\ra.$

The eternal black hole is described by the entangled state,
\be
|\Psi\ra = \sum_n e^{-\beta E_n/2} |n, n \ra
\label{thermofield state}
\ee
where $\beta$ is the inverse temperature of the black hole. The density matrix of each side is a pure thermal density matrix.

This state  can be interpreted in two ways. The first is that it represents the
thermofield description of a single black hole  in thermal equilibrium \cite{Israel:1976ur}.
In this context the evolution of the state is usually defined by a fictitious thermofield Hamiltonian which is the difference of Hamiltonians of the two CFTs.
\be
H_{tf} = H_R - H_L.
\label{thermofield hamiltonian}
\ee

The thermofield hamiltonian \nref{thermofield hamiltonian} generates boosts which are
translations of the usual hyperbolic angle $\omega.$  One can think of the boost as propagating upward on the right side of the Penrose diagram, and downward on the left.
The state \nref{thermofield state} is an eigenvector of $H_{tf} $ with eigenvalue zero, and is therefore boost invariant. The thermofield doubling of the Hilbert space and the introduction of $H_{tf} $ is a trick for facilitating the calculation of correlation functions for a system composed of a single copy. In this interpretation there is only one asymptotic region and one black hole in a thermal state.

The second interpretation of the eternal black hole is that it represents two black holes in disconnected spaces with a common time \cite{Maldacena:1998bw,Horowitz:1998xk,Balasubramanian:1998de,Maldacena:2001kr}.
We will refer to the disconnected spaces as sheets. The degrees of freedom of the two sheets do not interact but the black holes are highly entangled with an entanglement entropy equal to the Bekenstein Hawking entropy of either black hole.
We say that these black holes are ``maximally''  entangled\footnote{Note that the density matrix is the thermal density matrix and not the
identity matrix. By a slight abuse of language, we will still call these states ``maximally entangled''.}.

In this second interpretation the state \nref{thermofield state} is represents    two black holes at a particular instant of time  $t=0.$ In this interpretation,
the time evolution  is upward on both sides with Hamiltonian
\be
H = H_R + H_L.
\label{H}
\ee

The state \nref{thermofield state} is not an eigenvector of $H.$ Its evolution is given by,
\be
|\Psi(t)\ra = \sum_n e^{-\beta E_n/2} e^{-2iE_n t}|\bar n, n \ra.
\label{t-dependence}
\ee
where $|\bar n \rangle  $ is the CPT conjugate of the state $|n\rangle$. 

Although the state is {\it not}  time-translation invariant, the individual density matrices on either side are  time-independent thermal density matrices as before. Matrix elements of operators that depend only on one side
do not depend on time. Though the total entanglement does not depend on time, we will see that
more detailed properties of this entanglement do depend on time.
%To see the time dependence of the state we need to look at operators that perform simultaneous
%energy changing operations on the two sides. A correlation function of two local operators, one on
%each side, is an example.
% As another example, we can consider the operator,
%\be
%{\cal{O}} = |n,n\ra \la n+1,n+1 |
%\ee
%Its matrix elements evolve with a time-dependent phase factor $e^{2i(E_{n+1 } -E_n)t}$.
The time dependence is also evident from the Penrose diagram where one sees that the global geometry does not have an invariance under a time isometry that shifts both asymptotic times to the future.

The entanglement has a geometric manifestation. Even though the two black holes exist in separate non-interacting worlds, their geometry is connected by an \er.
The entanglement is represented by identifying the bifurcate horizons, and filling in the space-time with interior regions behind the horizons of the black holes.

% This connection between entanglement and Einstein-Rosen bridges will be central in what follows. In fact we believe it is a special case of a more general equivalence between entanglement and geometric connectivity.

\subsection{Cool horizons for entangled black holes}

The entangled  black holes described by the Penrose diagram in figure \ref{1}
 have no firewalls, and an observer who falls through the horizon does not feel anything special.
 But it is easy to change this. Let's expand  the system to include an observer Alice who lives on the left boundary of the Penrose diagram, and can control the boundary conditions. We can think of her as living asymptotically far away on the sheet  containing the left  black hole. Alice can send message into the bulk by manipulating the boundary condition on the left boundary.

Bob, on the other hand, lives in the bulk. He starts out on the right exterior region and may or may not cross the horizon. It it is obvious from the Penrose diagram that Alice cannot send a message to Bob as long as Bob does not cross the horizon of his black hole.

If Bob does cross the horizon he can receive a message from Alice if Alice sends it early enough (we postpone what early enough means until section \ref{Messages}). If Alice chooses, she can send a deadly message from a point very near the lower left corner of the diagram. For example she may shoot in a very high energy shock wave that will propagate upward to the right very close to Bob's horizon. This firewall  has no effect on anyone outside the horizon of Bob's black hole, but it kills anything that passes through the horizon. Its effects also decays exponentially as we move forwards in time along the horizon on the right.

Evidently the answer to the question---Does Bob's black hole have a firewall?---is that it depends on what Alice does.

One can also consider one-sided black holes in AdS. A one-sided black hole is modeled by a single copy of AdS. In this case, we can also send a shock wave as above, by sending in a shock wave in addition to the
infalling matter. The shock wave should be timed so that it does not escape from the black hole but lies just
inside the horizon.

%They are created in pure states by shooting in energy from the boundary. There does not seem to be any way to classically create firewalls in this case. A shock wave thrown in from the right boundary will just add to the mass of the black hole, but it will not appear as a firewall.

\subsection{Schwarzschild Black Holes}

\begin{figure}[h!]
\begin{center}
\includegraphics[scale=.3]{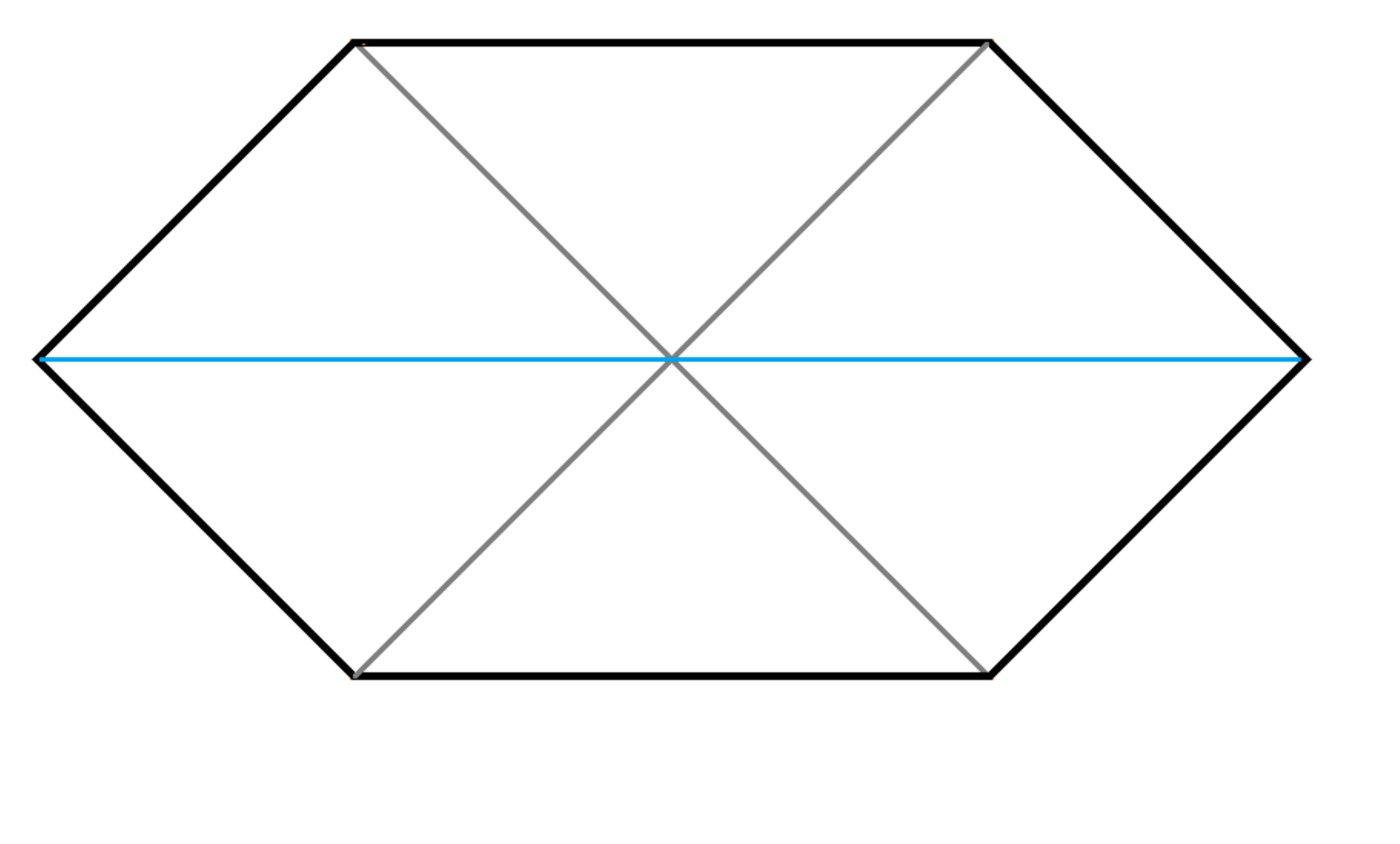}
\caption{Maximally extended Schwarzschild spacetime. There are two asymptotic regions. The blue spatial slice contains the \er\  connecting the two regions.   }
\label{2}
\end{center}
\end{figure}

\begin{figure}[h!]
\begin{center}
\includegraphics[scale=.28]{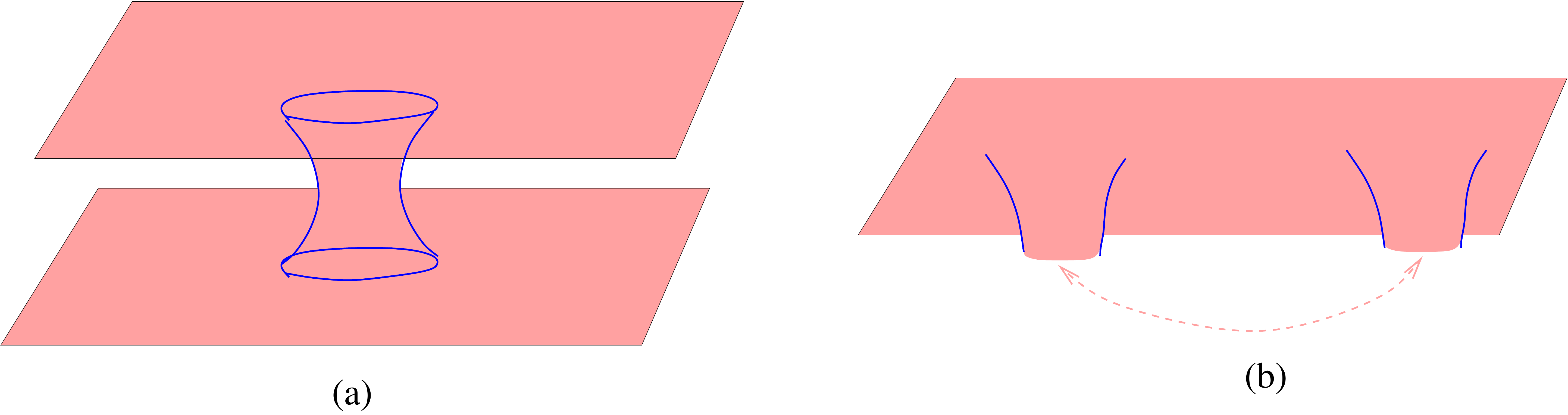}
\caption{(a) Another representation of the blue spatial slice of figure \ref{2}. It contains a neck connecting two asymptotically flat regions.
(b) Here we have two distant entangled  black holes in the same space. The horizons are identified as
indicated.
  This is not an exact solution of the equations but an approximate solution where we
can ignore the small force between the black holes.   }
\label{TwoSame}
\end{center}
\end{figure}

Entirely similar considerations apply to two-sided Schwarzschild black holes with Penrose diagram shown in figure \ref{2}. This time the two spatial sheets are asymptotically flat and each has an identical black hole. As before the black holes are maximally entangled but not interacting.
The blue line represents an initial instant at which the state has the form
\nref{thermofield state}. The entanglement is represented by the fact that the two bifurcate horizons touch at the origin. The spatial geometry on the $t=0$ slice looks like figure \ref{TwoSame}(a).
%   \begin{figure}[h!]
%\begin{center}
%\includegraphics[scale=.3]{3.pdf}
%\caption{ Another representation of the blue spatial slice of figure \ref{2}. It contains a neck connecting two asymptotically flat regions.  }
%\label{3}
%\end{center}
%\end{figure}

Although the geometry is connected through the bridge the two exterior geometries are not in causal contact and
information cannot be transmitted across the bridge. This can easily seen from the Penrose diagram, and is consistent with the fact that entanglement does not imply non-local signal propagation.

All of this is well known, but what may be less familiar is a third interpretation of the eternal \S \ black hole. Instead of black holes on two disconnected sheets, we can consider two very distant black holes in the same space. If the black holes were not entangled we would not connect them by a \er. But if they are somehow created at $t=0$ in the entangled state \nref{thermofield state}, then the bridge between them  represents the entanglement. See figure \ref{TwoSame}(b).  Of course,
 in this case,
  the dynamical decoupling is not exact,
  but if the black holes are far apart it is a good approximation. Note that the black holes in
  \ref{TwoSame}(b) can be separated by a large distance. But an observer just outside one horizon
  would be separated by a small spatial distance from an observer just outside the other horizon,
  at least at $t=0$.

We will imagine that Bob is stationed at  one black hole which we will consider to be the near black hole.  Alice is far away at the  far black hole. Near and far are of course interchangeable but we will look at the system through Bob's perspective. As long as Bob and Alice stay outside their respective black hole horizons, communication between them can only take place through the exterior space. This requires a long trip which cannot be short-circuited by the \er.

On the other hand, under certain conditions Bob and Alice can jump in to their respective black holes and meet very soon after.

Again in this case Alice can create a firewall on Bob's side if she throws in shock waves from her side early enough.

Finally we may consider one-sided black holes in flat space. These are just the ordinary black holes created by a collapsing system in a pure state.
 % There is no classical way to create a firewall in the one-sided case for the same reasons as in AdS.
However, we will see that  quantum theory allows one-sided black holes to eventually become two-sided, at the Page time. At the Page time, the emitted Hawking radiation carries as many degrees of freedom as the remaining black hole, and it is maximally entangled with the black hole. This early half of the Hawking radiation plays the role of the second black hole and as we will see, the question of firewalls in Bob's
 black hole will be decided by what Alice, who is very far away, decides to do with the radiation.

\subsection{Natural production of entangled black holes in the same spacetime}

The particular entangled state of two black holes,   that we have been discussing,
 is very special and
one might worry that it would be extremely difficult to produce.
Here we point out that the process of black hole pair creation in an magnetic (or electric)
field \cite{Garfinkle:1990eq} is such that the pair is precisely in this state.

\begin{figure}[h!]
\begin{center}
\includegraphics[scale=.35]{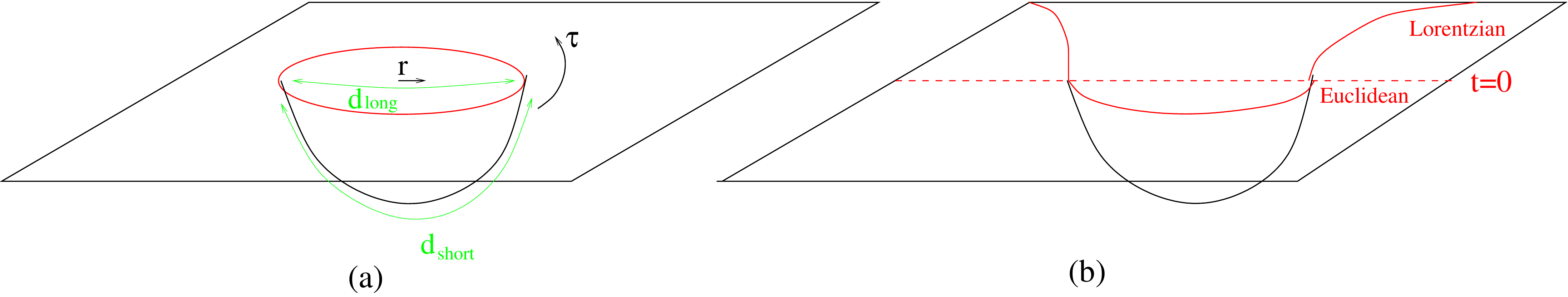}
\caption{ (a) Picture of the  Euclidean
 instanton describing  the creation of a black hole pair in a magnetic field. The space consists of two pieces joined along a cylinder with the topology $S^1 \times S^2$. The
$S^1$ is the circle drawn here. The bottom ``cup'' represents an approximately $H_2 \times S^2 $
geometry which is
the near horizon geometry of the extremal black holes. Despite appearances the distance through the
$H_2 $ section is shorter than through the plane. (b) Lorentzian continuation across $t=0$ gives a
pair of accelerating black holes.  }
\label{PairCreation}
\end{center}
\end{figure}

One considers a geometry with a constant magnetic field. The pair of black holes is
described by a certain Euclidean instanton geometry, see figure \ref{PairCreation}.
There is a one parameter family of instantons
that describe the creation of a pair of black holes of various sizes. The dominant
case corresponds to the creation of relatively small black holes. These black holes are close to
extremality. Extremal black holes have a fixed charge to mass ratio. This means that their acceleration
in the presence of a magnetic field is set by the magnetic field and is independent of their mass.
The Euclidean instanton contains a charged
 black hole going around a circle in Euclidean time. The radius
of this circle is set by the acceleration. This acceleration also leads to a Rindler temperature. The
black holes are at this temperature. They are in equilibrium with the bath of radiation.
Interestingly, the total pair creation rate has an overall factor of $e^S$, where $S$ is the entropy
of one of the black holes
\cite{Garfinkle:1993xk}. This is precisely as expected if the two black holes are in this entangled
state. The metric of the instanton is written in appendix A, as well as its approximate form for
small black holes.

\subsection{Different bridges for different entangled states}
\label{difffordiff}

As it is well known, entanglement is not an observable quantity in the sense of Dirac.
Namely, there is no projection operators $P_E$ which acts on two Hilbert space factors $H_A$ and $H_B$
such that $P_E \not =0$ for  entangled states and $P_E =0$ for unentangled states.
The reason is simple,
we can write an entangled state as a linear combination of unentangled states. Thus no such
operator exists.
Note, however, that we can define projection operators to particular entangled states. For example,
we can ask whether two qubits are in the total spin zero state.
We can also ask whether two qubits are in the spin one,  $J_z=0$,  state. These
are two entangled states
\be
|\Psi_0 \rangle  \sim  |+ \rangle |- \rangle - |- \rangle | + \rangle ~,~~~~~~~~~~
|\Psi_1 \rangle  \sim |+ \rangle |- \rangle + |- \rangle | + \rangle
\ee
There is a standard projector operator $P_0 = |\Psi_0 \rangle \langle \Psi_0 | $ that
tests whether the system is the entangled state $|\Psi_0\rangle$, and a different one
that tests whether it is in the other state.

We claim that we should think of the  bridges associated to these two states
 as being different.
In fact, we can see this clearly in the case of the eternal black hole.
In this case,  we can consider the following family of  Schrodinger picture states
\be \label{familyt}
% |\Psi_0 \rangle \sim \sum_n e^{ - \beta E_n/2} |n,n\rangle ~,~~~~~~~~~~{\rm or} ~~~~~
 |\Psi_t \rangle \sim \sum_n e^{ - \beta E_n/2} e^{ - 2 i E_n t }  |n,n\rangle
\ee
Two states with different values of $t$   are related by forward time evolution on the two sides. However, consider them as possible alternative states at the same instant of time and view $t$  in
\nref{familyt} as a parameter labeling alternative states at a common instant of time.
All these states have ``maximal'' entanglement and the same
density matrix on each side.  There is a projection operator $P_t$ into
each of these states. However, there is no projection operator onto the whole family, since
considering linear combinations such as $\int dt e^{ 2 i E_0 t} |\Psi_t \rangle$ projects us into
a particular state $|n_0,n_0 \rangle$,  which is the one having the energy $E_0$. This state is not
maximally entangled.

 \begin{figure}[h!]
\begin{center}
\includegraphics[scale=.5]{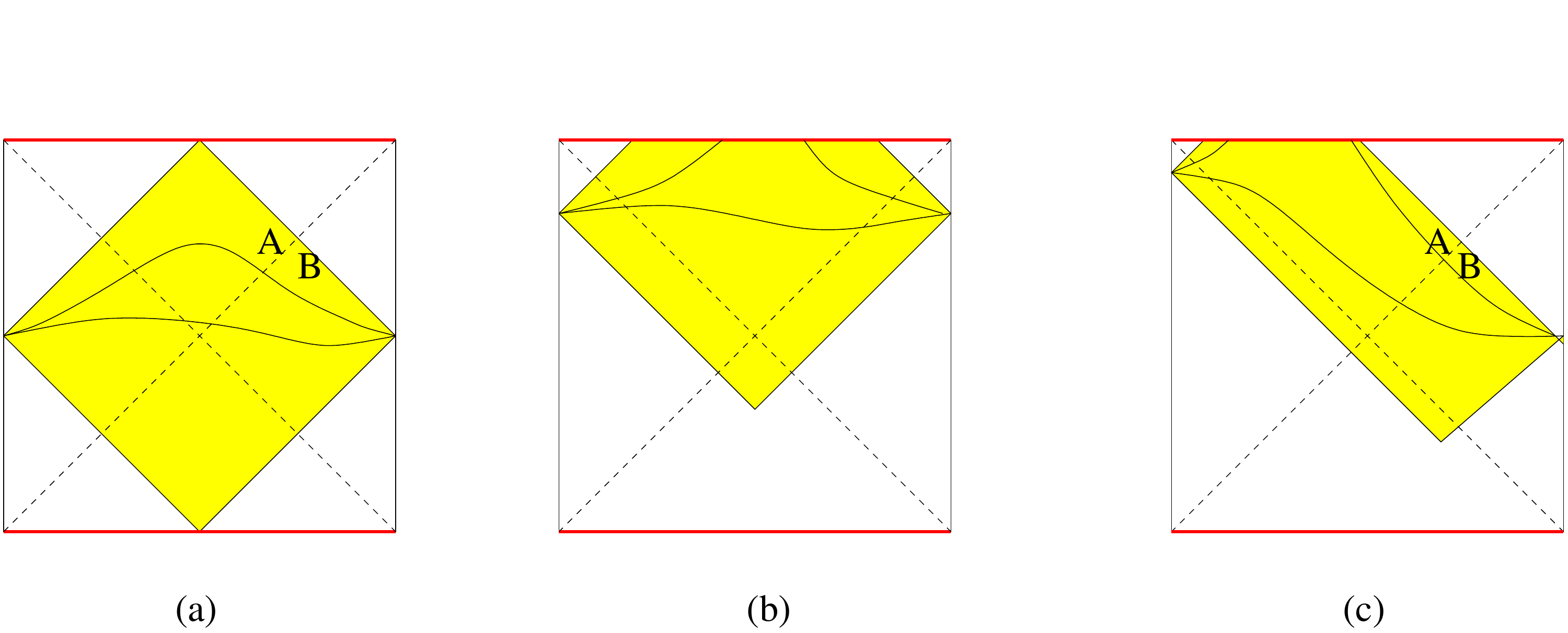}
\caption{  (a) The yellow shaded region corresponds to the Einstein Rosen bridge
associated to the entangled state $|\Psi_{t=0}\rangle$ in  \nref{familyt} (for an
$AdS_3/CFT_2$ situation).
 One can draw
different spatial sections in the geometry. The physics in these slices is related by the bulk
Wheeler deWitt equation. (b) Here we see the bridge corresponding to the entangled
state $|\Psi_t\rangle$, for $r>0 $. (c) This is a different presentation of the same
bridge as in (b), related by the action of a boost $H_R - H_L$.  Even though the
states (a) and $(c)$ are different, they both contain regions $A$ and $B$ which look the same.
   }
\label{DifferentBridges}
\end{center}
\end{figure}

Now, we claim that the precise bridge  associated to each state in the family \nref{familyt}
corresponds to a different   bridge. For each particular $t$ we want to assign a
region of the bulk. Since there is no preferred time slicing in the bulk, we can just simply assign
the whole spacelike separated region in the bulk. Namely, all points in the bulk that are
spacelike separated from the chosen boundary time. Different spatial slices in this region
are linked by
the bulk Wheeler deWitt equation. This seems a reasonable choice, at least from semiclassical
bulk considerations, see also \cite{Heemskerk:2012mn}.
As we see in figure \ref{DifferentBridges}, the bulk regions, defined in this way are all different.
 For example,
 in \ref{DifferentBridges}(a) we see that for the $|\Psi_0\rangle$ bridge
  we do not include a segment of the spacelike singularity, while at $t> 0$, in figure  \nref{DifferentBridges}(b,c),  we
include a segment of the spacelike singularity.

In conclusion, we have different bridges for different entangled states.
Note that the density matrix for each copy of the field theory is identical for all these states.
The only difference between the states is in the nature of the entanglement, in the precise unitary
transformation that relates them.

\begin{figure}[h!]
\begin{center}
\includegraphics[scale=.5]{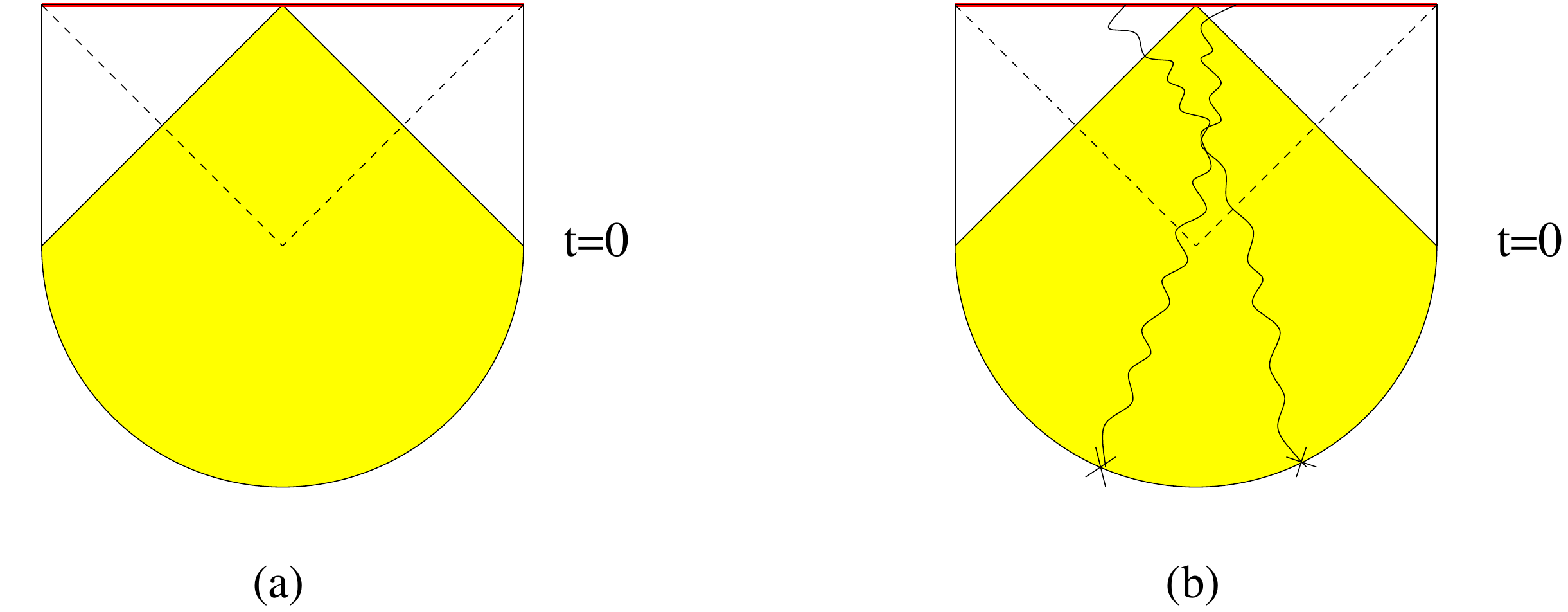}
\caption{ (a) Construction of the entangled Hartle Hawking state from Euclidean
evolution. This creates very particular entangled bulk and boundary states. (b) We can
add particles on top of the Hartle Hawking vacuum by adding operators on the boundary theory.    }
\label{WithParticles}
\end{center}
\end{figure}

So far, we have considered a very special family of unitary transformations. Namely, the same family
that is generated under time evolution. But we can easily consider other possibilities.
For example, suppose that we have the entangled black holes described by $|\Psi_0\rangle$.
This state can be produced by performing Euclidean evolution over an amount $\beta/2$ of Euclidean time. See figure \ref{WithParticles}(a). This prescription can be used both in the bulk and the boundary
theories. In the bulk, it produces the usual Hartle Hawking state for the quantum fields propagating
on the black hole background.
Two modes, one on each side of the horizon, are entangled with very particular phases in this state.
If we were to change these phases we would be producing particles in the bulk\footnote{Provided we
change for a small number of modes and in a suitably smooth way we get a small number of particles
in the bulk. We restrict to this case here. }.
%  If we do this in a suitably smooth way, and for a small number of modes, the new
%state will be the usual Hartle Hawking state with some additional particles.
We can give a rather precise construction of this state.
It can be obtained
by doing time evolution with operators inserted as in figure \ref{WithParticles}(b).
The precise formula
relating local operator insertions and the bulk wavefunctions of the corresponding states is given in
equation (2.7)  of  \cite{Maldacena:2001kr}.
 In general, adding particles to the Hartle Hawking state will modify the
density matrices of the either side.

 However, if the particles were produced by simply acting with a unitary transformation on the left side, then the density matrix on the right side will not be changed.

 These are particular cases of the general picture in  figure \ref{WithParticles}(b). They are special because    the
inserted operator along the Euclidean boundary is acting as a unitary transformation.
The final entangled state will differ from $|\Psi_0 \rangle$ by a  unitary transformation
acing on one of the copies of the field theory.
Depending on the precise unitary transformation we will get states with different particles.
These are further examples of different entangled states corresponding to different bridges
between the two sides.

Of course, we can also consider bridges, as in figure \ref{WithParticles}(b), with an
arbitrary configuration of bulk particles. These are all different bridges corresponding
to different entangled states, though they are not all maximally entangled.

\subsection{Bridges for less than maximal entanglement}

\begin{figure}[h!]
\begin{center}
\includegraphics[scale=.5]{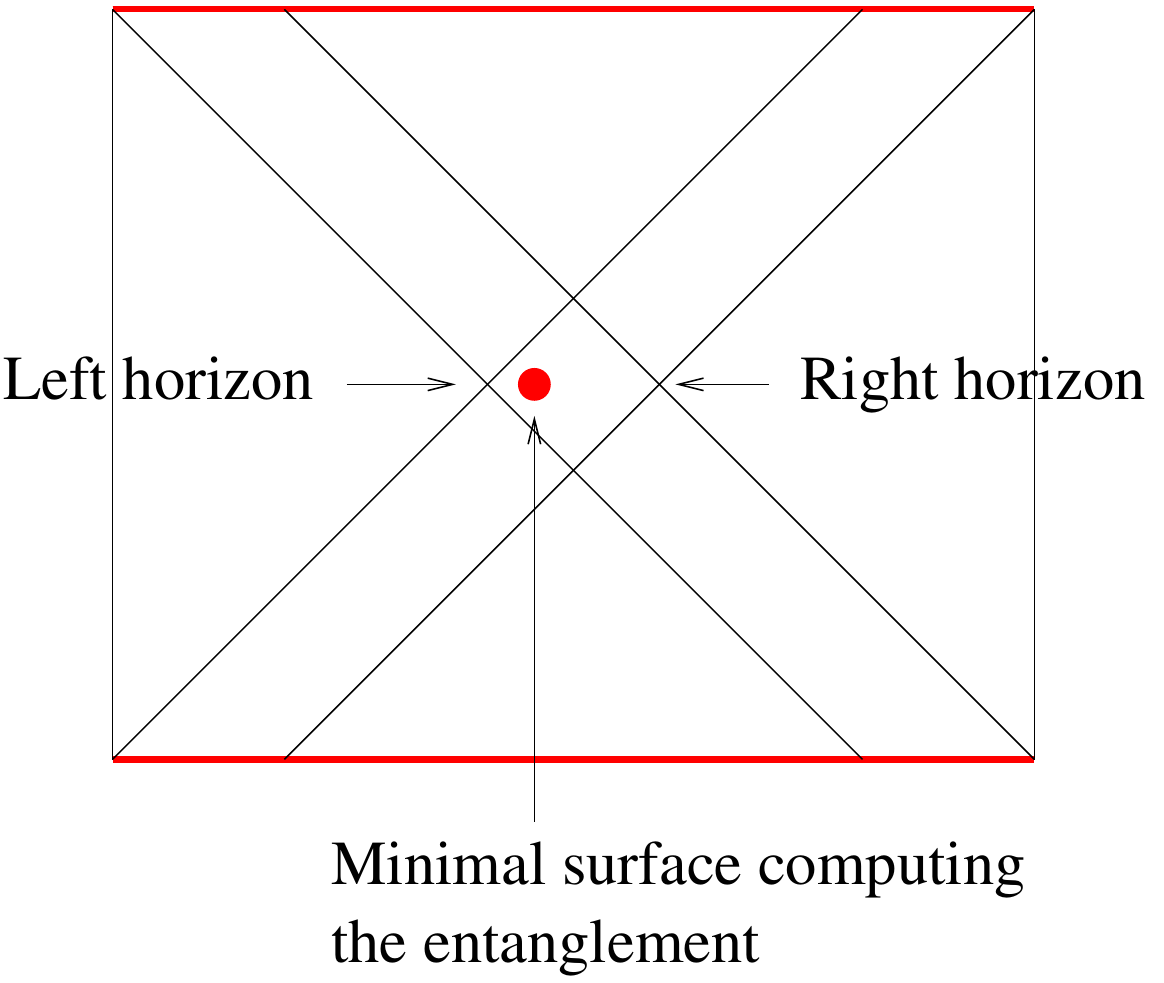}
\caption{ Penrose diagram of a configuration obtained by analytic continuation of a
time reflection symmetric, but  time dependent,  Euclidean solution.
The two horizons do not touch. The
entanglement, computed by the Ryu-Takayanagi prescription \cite{Ryu:2006bv},  is given by the area of
 a minimal surface with
less area than the horizons.The area of the horizons grows when we go from the bifurcation point to the future.   }
\label{NonMaximal}
\end{center}
\end{figure}

In the Penrose diagrams we have discussed the Left and Right horizons touch each other.
It is also possible to have configurations where they do not touch each other.
A simple way to generate them is to start from two eternal black holes and add some matter to
each side. These configurations can also be prepared by considering Euclidean evolution
with a time dependent Hamiltonian,  see  \cite{Bak:2007qw} for some explicit
solutions\footnote{ The solutions in \cite{Bak:2007qw} are based on Janus solutions. Their
   boundary in Euclidean space has the form $S^1 \times \Sigma$ where $\Sigma$ is
  a quotient of hyperbolic space. The $S^1$ is divided in two equal parts and the dilaton has
  a different value on each part. The Lorentzian continuation is obtained by continuing across the
  moment with a time reflection symmetry. The two boundaries different values for the dilaton. These values
  are constant in time. The bulk smoothly interpolates between the two.  }.
The Penrose diagram of such configurations is given
in figure \ref{NonMaximal}.

% \section{Geometry and Entanglement}

\subsection{Growth of the Bridge}

\begin{figure}[h!]
\begin{center}
\includegraphics[scale=.3]{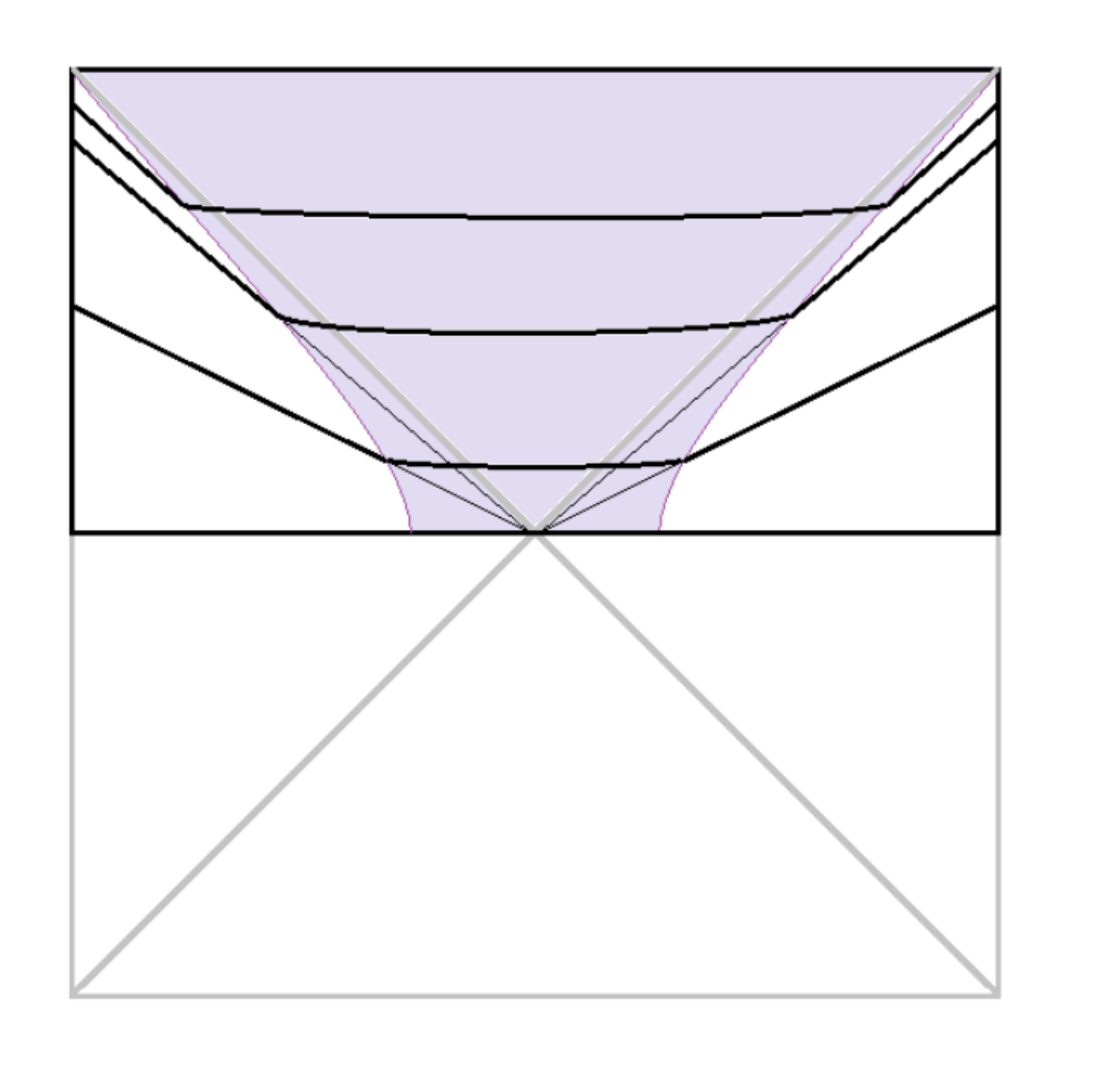}
\caption{ Equal time slices for the eternal black hole. The slices grow in size in the
interior. The spatial distance between opposite points on the stretched horizon grows.   }
\label{4}
\end{center}
\end{figure}

 \ers \ are not traversable.
 As an observer jumps into the black hole, he sees that the transverse two sphere shrinks as he
 approaches the singularity. Thus, it is sometimes said that the bridge closes, or pinches off, before
 he can get through \cite{Fuller:1962zza}.
  A related feature is  that as global time evolves the bridge stretches. Its length grows so fast that no signal can get through.
  % The precise statement depends on the slicing used to define global time.
  In figure \ref{4}
a particular slicing of the upper half of the Penrose diagram is indicated. A stretched-horizon is introduced on each side and the slices are conventional time-slices outside the horizons. Inside they are somewhat arbitrary. An invariant statement is that the spatial distance between a point on the L horizon and one on the R horizon grows as we move these points to the future, keeping them on the horizon.

There seems to be an intimate connection between the entanglement of the underlying degrees of
freedom and the geometry of spacetime. A clear manifestation of this is the Ryu-Takayanagi formula
for entanglement entropy in the gauge/gravity duality \cite{Ryu:2006bv}.   Van Raamsdonk \cite{VanRaamsdonk:2010pw} has also conjectured that the amount of entanglement between two regions
is related to their distance. The greater the entanglement the less the distance. 

Time evolution is the simple innocent looking operation of adding the phases in
\nref{t-dependence}. We will now discuss how the entanglement grows under this operation,
see \cite{Hartman:2013qma} for further discussion.

\begin{figure}[h!]
\begin{center}
\includegraphics[scale=.2]{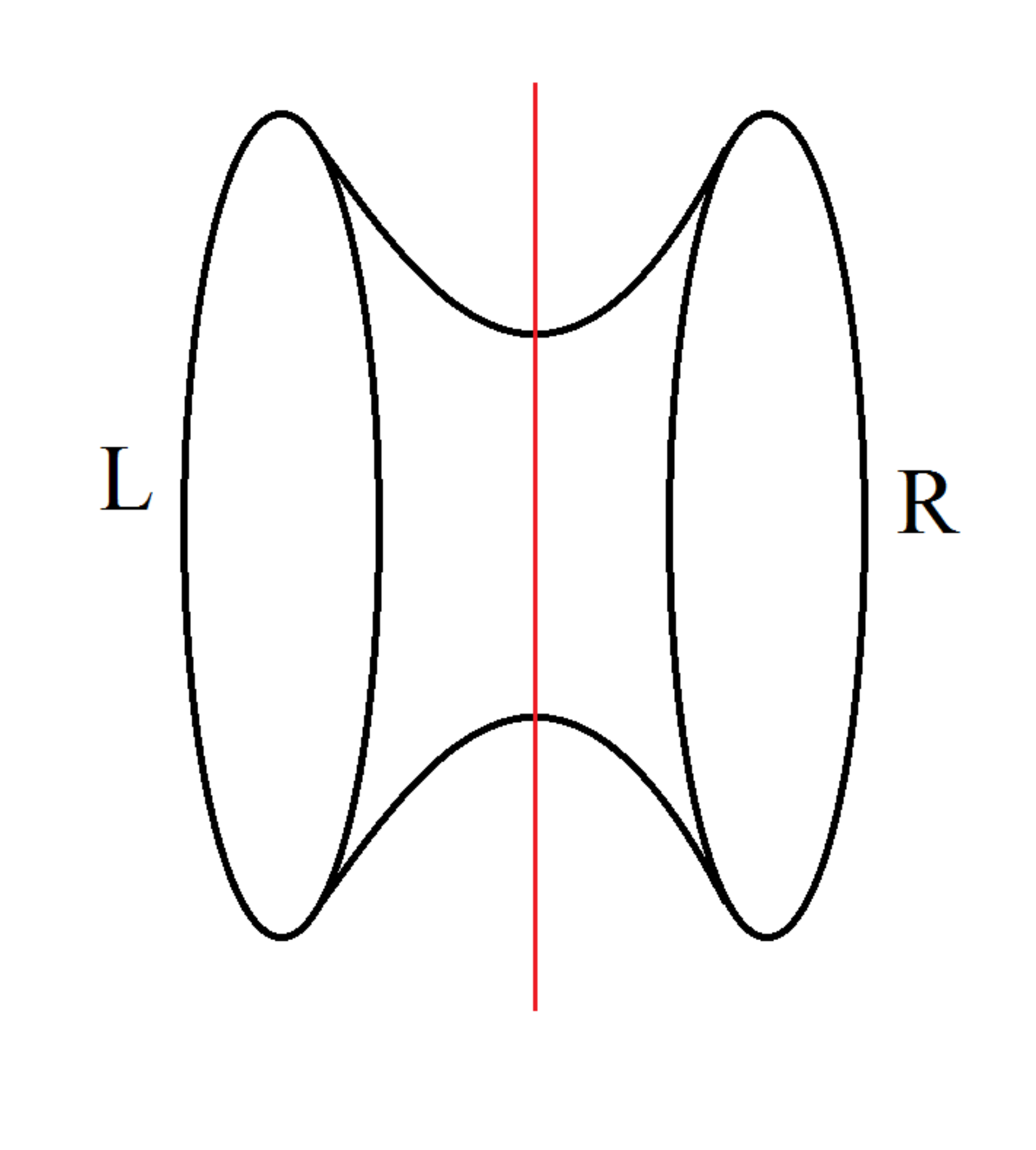} ~~~~~~~~~~~~~\includegraphics[scale=.2]{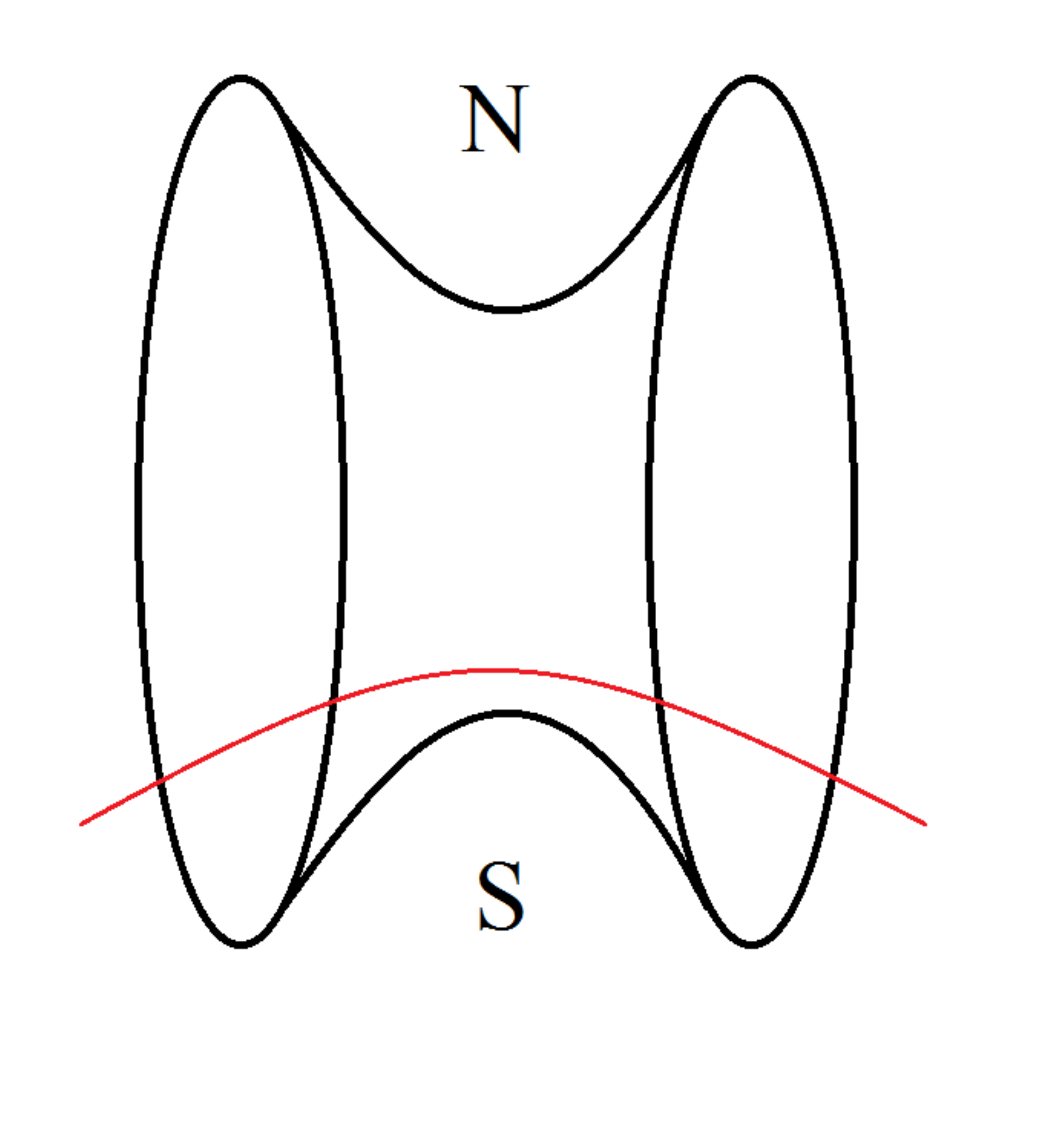}
\caption{ We consider a pair of entangled black holes with spherical horizons. We call them L and R.
They have a bridge joining them. We divide them into two parts in two ways. The vertical line
separate the left and right degrees of freedom. We can also produce a cut that separates both spheres
into a northern and southern parts.    }
\label{5}
\end{center}
\end{figure}

In what follows we will divide the spatial  geometry in various ways. First  we  divide it in the standard way into  left and right halves $L, \ R,$ (corresponding to CFTs L and R shown in figure \ref{5}).

 We will also make a second division into northern and southern regions, $N, \ S,$ as in figure \ref{5}. The division is not equal. The southern region is smaller than the northern.

 Finally we can divide the geometry into four sectors, $NL, \ NR, \ SL, \ SR.$

We will model the \dof \ of these regions by a system of $2N$ qubits: $N$ for the left side and $N$ for the right side. We can further divide them among the four subsystems,  $NL, \ NR, \ SL, \ SR.$

From the properties of the Hartle Hawking state the Left and Right qubits are initially entangled in a state which is a tensor product of Bell pairs. The qubits are localized on the sphere and the entanglement is between qubits at similar location. This means that there is very little entanglement between north and south. There is however a maximal entanglement between $NL$ and $NR$ and also between $SL$ and $SR.$

%   \begin{figure}[h!]
%\begin{center}
%\includegraphics[scale=.3]{6.pdf}
%\caption{  }
%\label{6}
%\end{center}
%\end{figure}

%   \begin{figure}[h!]
%\begin{center}
%\includegraphics[scale=.3]{7.pdf}
%\caption{  }
%\label{7}
%\end{center}
%\end{figure}

Now we allow the system to evolve by means of the Hamiltonian $H_1+H_2.$   The time development is a product of left and right evolution
\be
U = U_1 U_2
\ee
We expect that dynamics of each side is chaotic,  and that it scrambles in a Rindler time of order $\log{S}. $
After the scrambling time the operators $U$ are represented as random matrices in the qubit system. The effect is to scramble the entire system.
We  expect the following features:

\bi
\item The total entanglement entropy  between the right and left side has not changed.
% By Page's argument
It is still maximal.
\item The total north-south entanglement increases to its maximal value, which
is equal to the number of qubits in the smaller southern region.
 This increase is directly related to the growth of the bridge mentioned earlier, \cite{Hartman:2013qma}
(see also \cite{Morrison:2012iz}).
% This is the entanglement of (NL+NR) with (SL + SR).
\item The lack of entanglement between $NL$ and $SL$ remains unchanged.
 %That is because they are small subsystems (less than half the number of degrees of freedom)
\item The initial large entanglement, as well as the mutual information between $SL$ and $SR$ tends to zero,   because these are small subsystems (we take the South to be smaller than the North).
    Van Raamsdonk's argument  suggests that
     the distance between the left and right region grows.

\ei

\sc
\section{ ER= EPR}

There are similarities between entanglement (EPR) and Einstein-Rosen Bridges that we want to call attention to.
In fact,  we are going to take the radical position that in a theory of quantum gravity they are inseparably linked, even for systems consisting of no more than a pair of entangled particles.

\subsection{No Superluminal Signals}
At first sight both entanglement and \ers \ seem like strange violations of locality,  but in both cases
they do not provide mechanisms for superluminal signal propagation. It is easy to prove that no local operation on one member of a
entangled pair can influence the other,  before a classical signal can propagate between them.

The situation is the same for \ers. A look at any Penrose diagram of a two-sided black hole easily shows that no signal can propagate
through the wormhole from one exterior region to the other.

\subsection{No Creation By LOCC}
In quantum information theory LOCC stands for local operations and classical communication. It refers to two separated systems---call them Alice's and Bob's shares---
and the possible operations that \it do not \rm increase the entanglement between them. An LO operation is any quantum or classical operation that Alice and Bob
can do on their own shares. This could include product unitary operations $U_A\otimes U_B,$ local measurements, or processes involving ancillary systems.  The CC of LOCC stands for classical communication.
%Classical communication means exchanging information in a form that does not involve non-commuting operators.
CC allows Alice to send Bob a message reporting the result of a measurement of the $z$-component of spin.
What she  cannot do is to send the quantum spin itself.
%, or to quantum teleport it.
It is a basic tenet of quantum information theory that
LOCC cannot create or increase entanglement.

 There are two ways to create entanglement between Alice's and Bob's shares. The first is to bring them together into direct contact and allow them to interact by quantum-mechanical interactions.
 The second is to have a resource of entangled qubits, which of course had to be prepared in advance by direct interactions. If Charlie creates a collection of Bell pairs, and sends half of each pair to Alice, and the other half to Bob, they can use
 the resource to entangle their shares.

 The situation for \ers \ appears to be very similar. Given two distant black holes with no \er, there does not seem to be any way to create a bridge between them without preexisting bridges. However, it is possible to create a neighboring black hole pair connected by an \er,  and then separate the pair.
  %A known example is the pair creation of near-extremal black holes in an electric or magnetic field. The geometry of the instanton contains a wormhole between the pair.  The field then accelerates the pair away form one another. The analysis of the process shows that the probability contains a factor $e^S$ where $S$ is the entropy of one member of the pair. This implies that the pair is maximally entangled.
 %The horizons of the black holes are Rindler-like. This is because the black holes created in this way are not at zero temperature. Since they are accelerating in the electric field they are in equilibrium with Unruh radiation.

 One can also consider a pair of maximally entangled extreme black holes at zero temperature. Since all the states of an extremal black hole are degenerate, the phase factors in \nref{thermofield state} are all identical. Thus the state evolves with a trivial overall phase factor that has no physical significance. This leads to a puzzle: if the state does not evolve then it seems that the \er \ does not grow with time, and it would be traversable. The resolution is that the \er \ of an extremal black hole has infinite length to begin with. This is just the statement that the length of the throat between the exterior and the horizon of a charged black hole tends to infinity  as the temperature tends to zero. We point out that the area of the bridge stays finite, i.e., it does not pinch off.

 The Penrose diagram describing an extremal black hole does not have two sides connected by a bridge. It is not a good description of the two entangled black holes. A better way to describe them is by using the non-extremal geometry choosing the temperature to be below the gap to the first excited state
 \cite{Sen:2011cn}.

 Another way to create a bridge between two separated black holes is to use a second pair that already has a bridge. Then by merging the members of the original pair with the new pair, a bridge will be formed.
 We can even go to a suggestive extreme: create a large number of small black-hole-pairs with bridges. Give half of each pair to Alice and the other half to Bob and allow them to separate to a large distance. Then let Bob and Alice  each merge their own shares into a single black hole on each side. The two resulting black holes will be connected by a bridge.
 In these cases the black holes might have less than maximal entanglement, then the
  bridge's neck might have an area smaller than the area of the black hole horizon.

 Finally let us make a jump. Suppose that we take a large number of particles, entangled into separate Bell pairs, and separate them in the same way as the mini-black holes.
When we collapse each side to form two distant black holes, the two  black holes will be entangled. We make the conjecture that they will  also be connected by an \er. In fact we go even further
 and claim that even for an entangled pair of particles, in a quantum theory of gravity there must be a Planckian bridge between them, albeit a very quantum-mechanical bridge which probably cannot be described by classical geometry.

 We summarize out conjecture with the symbolic equation,
\be
ER = EPR
 \ee
 and suggest that it is a completely general relation.

 \cite{Marolf:2012xe,Marolf:2008tx} proposed that one needs to
introduce  superselection sectors in order to have non-trivial topologies.
According to the $ER=EPR$ connection, such non-trivial topologies
simply characterize the entanglement between the black holes and should be allowed as possible
quantum states.

 \subsection{ Restoring the thermofield state}

We can have a pair of maximally entangled  black holes which are not in the thermofield state
\nref{thermofield state}. By assumption,
 the density matrix for each of the two black holes  is the thermal density
matrix.
%
%
%Distant black hole pairs made by these methods  can be maximally entangled, but the wave function is  unlikely to have the real phases that we associate
%with the thermofield state \nref{thermofield state}. But without violating any principle of quantum %operations. Let us suppose the state comes out in the form,
%\nref{simpleTF} except with arbitrary phases.
%\be
%|\Psi\ra = \sum_n F(n) e^{i\phi_n} |n,n\ra
%\label{arbphase}
%\ee
%
We will assume that Alice has a very powerful quantum computer which is not limited by any technological constraints. It can perform arbitrarily complicated quantum-computations in an arbitrarily short amount of time. In this respect we are following the assumptions of AMPS and ignoring possible limits of computational complexity \cite{Harlow:2013tf}.
%We also assume that
% Alice knows the dynamics well enough to know the phases $\phi_n.$
 Equipped with those powers, Alice  can put her black hole into her quantum computer which she has programmed to act with a unitary
operator $U_A$ in the Hilbert space of her share. This unitary operator in principle can set the
state to the thermofield state \nref{thermofield state}.
% The Unitary is given by
%\be
%U_A = \sum_n |n\ra \la n| e^{-i\phi_n}.
%\ee
%
%The outcome of the operation is to readjust the phases in \nref{arbphase} to zero.
 Note that the operation does not change the entanglement entropy between Alice's share and Bob's.
 If we had started with the thermofield state and had let it evolve so that it acquires the
 phases \nref{t-dependence}, then Alice could ``rejuvenate'' it by applying the unitary operator
 that reverses these phases.

\subsection{Messages From Alice  to Bob }

\label{Messages}

\begin{figure}[h!]
\begin{center}
\includegraphics[scale=.6]{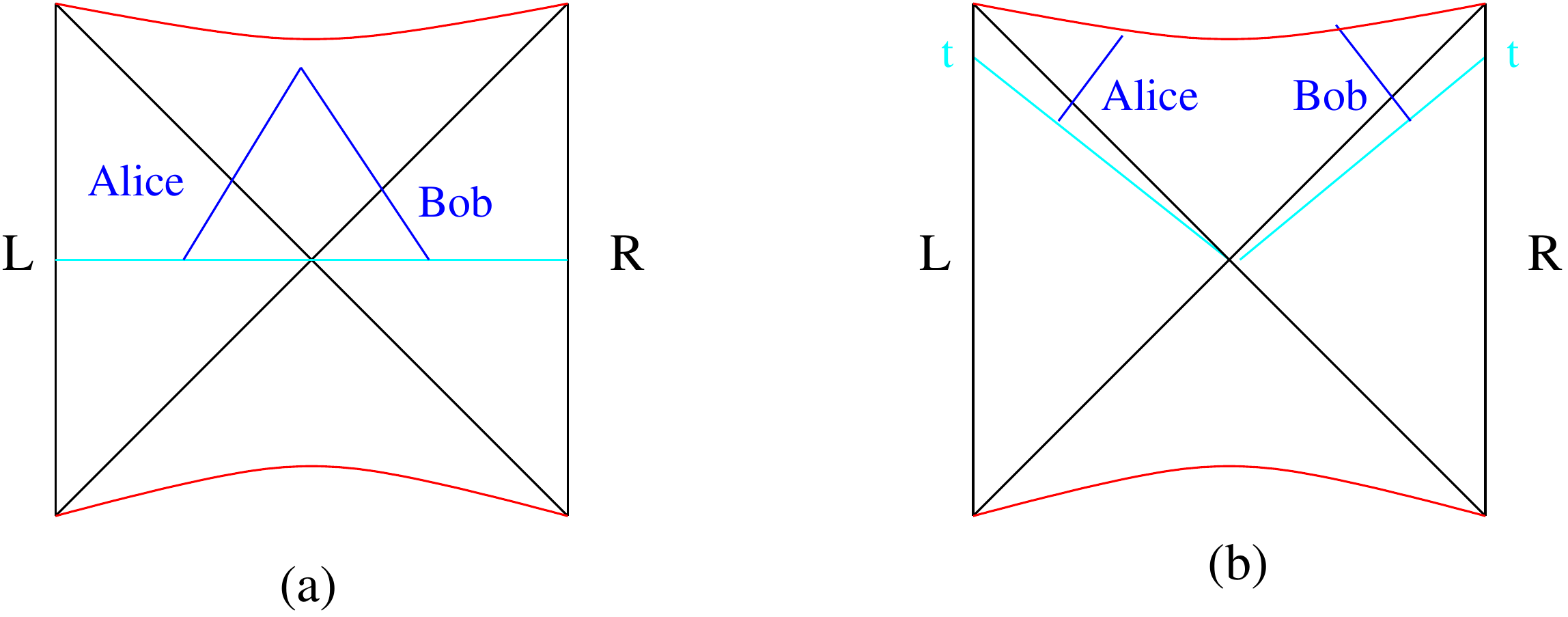}
\caption{Two black holes in the entangled thermofield state. (a) If Alice and Bob each jump into their
respective black holes, they can meet in the interior. (b)  If they wait too much
 they will not  meet.   }
\label{11}
\end{center}
\end{figure}

Returning to the case of a near and a far black hole, we know that it is impossible for Alice to send Bob a superluminal through the bridge, as long as Bob stays outside the horizon of the near black hole.
 If Bob jumps into the black hole he can get Alice's message.
 However, if they wait too long,   it may be too late for her to send a message that intercepts Bob even after he passes the horizon, see  figure \ref{11}(b).

But Alice has a plan; she uses her quantum computer to reset the phases on her side so that the phases in the wavefunction become real. This effectively runs her time back to $-t$. Then she jumps in and easily meets Bob.
Alice could also send in
  dangerous messages including deadly firewalls. But Bob has no way of knowing what Alice will do until he passes the horizon. He may get flowers or he may get bullets.

The lesson seems to be that we can neither be sure whether there is a firewall even though the conditions for the AMPS argument are satisfied. It all depends on what Alice does.

 \subsection{Clouds}

\begin{figure}[h!]
\begin{center}
\includegraphics[scale=.3]{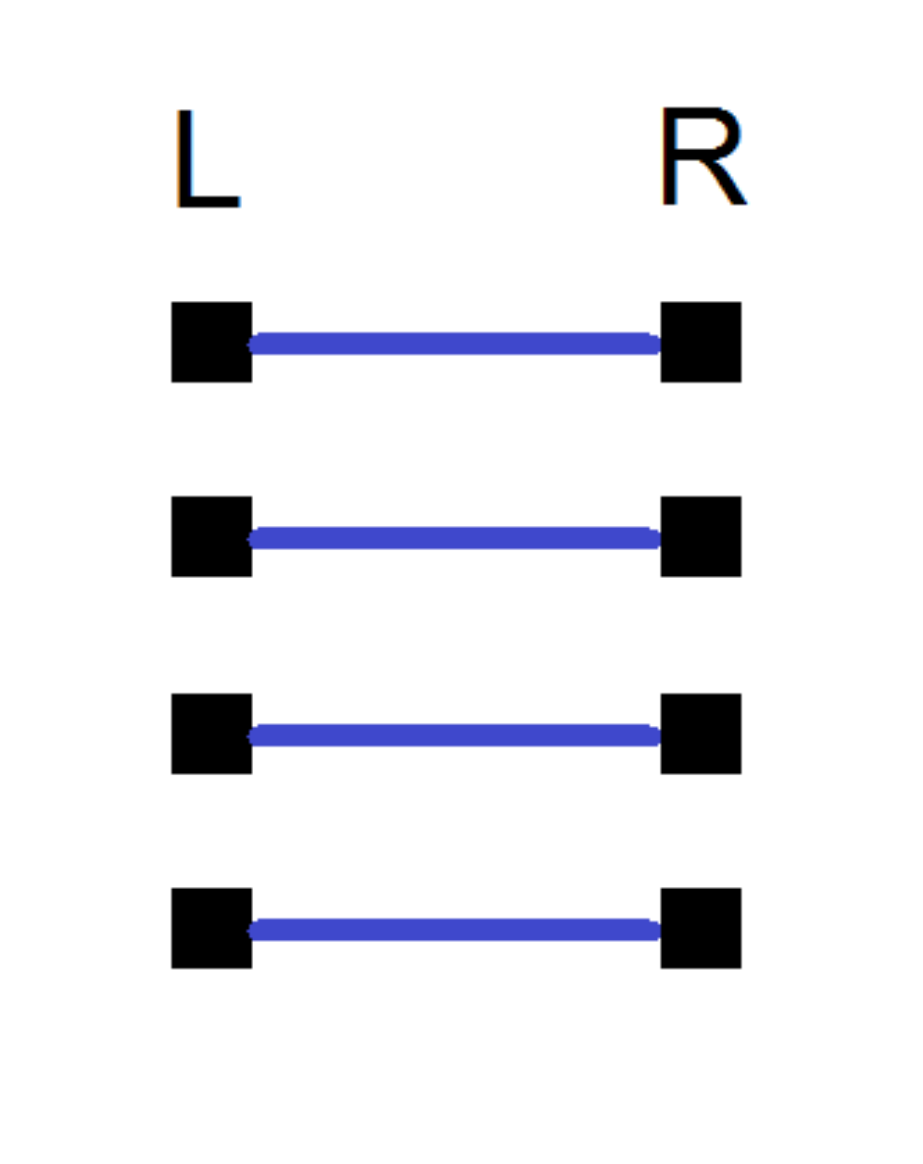}
\caption{ Entanglement pattern for a collection of Bell pairs.   }
\label{8}
\end{center}
\end{figure}

 Black holes are not the only things that can be entangled. To understand $ER=EPR$ better let us
 consider a cloud of $2N$ entangled particles (qubits). It is not important how they are distributed in space, but it is very important what the pattern of entanglement is. Assume the qubits are distributed into two shares.
 In the first illustrative case  the qubits are entangled in Bell pairs, each of which is split into Alice's share and Bob's share. The state can be written
\be
 \Psi = |singlet\ra^{\otimes N}
 \ee
The \ers \ connect the particles in pairs, schematically illustrated in figure \ref{8}.
Note that there is maximal entanglement between Bob's share and Alice's share, but if we divide the system into to shares differently, there may be no entanglement.

 Now let us  scramble the state by applying a random unitary operator to it.
 % It's enough to apply the unitary to one side.
  The system will become scrambled, and there will be maximal entanglement between any two subsystems, no matter how they are divided up. Once scrambled the system cannot be divided into unentangled subsystems. This means that the system of bridges in figure \ref{8}
cannot correctly describe the state. It is clear that the system of \ers \ in the scrambled  case must be connected.

\begin{figure}[h!]
\begin{center}
\includegraphics[scale=.3]{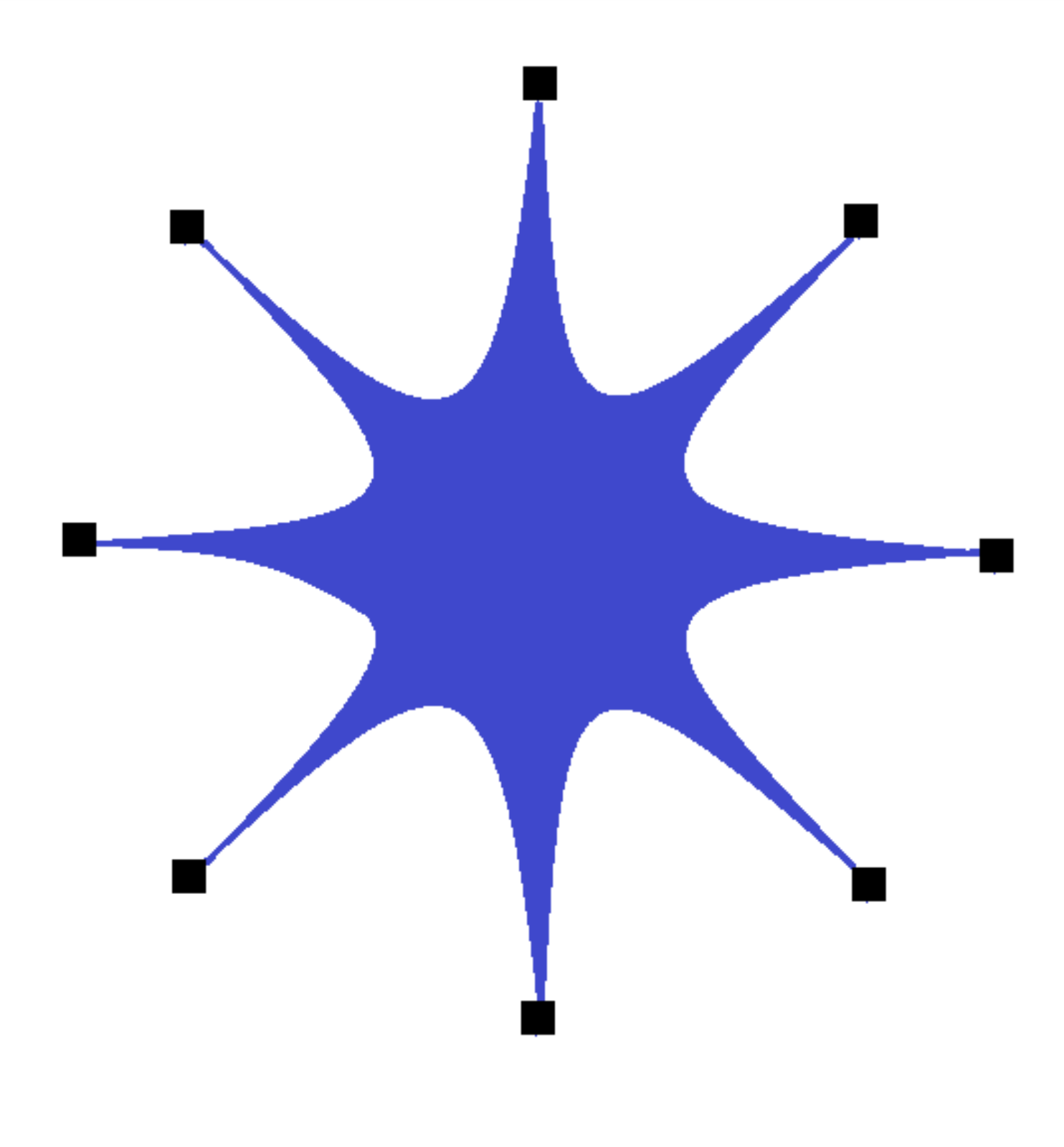}
\caption{ Entanglement pattern for a generic state of qubits. The black dots are the qubits and
the shaded region attempts to capture the pattern of entanglement. }
\label{9}
\end{center}
\end{figure}

There are a huge number of such states, since the typical state in Hilbert space if of this form.
% (, of order $e^{2^{2N}}$This is the volume of the
%Hilbert space of $2N$ qubits.)
 It is likely that  no simple four-dimensional classical geometry exists for most of them. Possibly, among them there are some for which the \er \  can be interpreted geometrically. The natural guess for these would be a large ``nucleus" with ``legs" connecting the nucleus to the qubits in the external space,
 see fig. \ref{9}.
 % Schematically, this looks  like figure \ref{9}.

We know very little about such configurations, but there is one constraint that should be satisfied, assuming the nucleus has a geometry that can be described by Einstein's equations. Consider cutting through the \er \ and dividing it into two pieces. Suppose one piece contains $N_a$ legs and the other $N_b$ legs with $N_a \geq N_b$. The entanglement entropy between the subsystems is the number of qubits in the smaller system, $N_b $  \cite{Page:1993wv}.

The cut through the \er \ defines a two dimensional surface whose area should not be smaller than the entanglement entropy. Based on  \cite{Ryu:2006bv}  we expect
 that the smallest area of such a cut is, in fact, the entanglement entropy.
Thus the area of the cut should be of order $N_b.$ Overall, the size of the nucleus should be of order the number of qubits.\footnote{ As a less entangled state we could consider the ground state of a
spin system that has an IR fixed point. The pattern of entanglement is governed by the renormalization group. This pattern is reminiscent of hyperbolic space  \cite{Swingle:2012wq}. In a system with a gravity dual the bridge
is the spatial section of $AdS$.}

Now imagine collapsing the cloud into a single black hole in a pure state. In either case---figure \ref{8} or \ref{9}---the system will become scrambled. The qubits will collect to form the stretched-horizon and zone of the black hole. Whether or not they were initially scrambled, after a time of order $M\log M$   they will become scrambled and therefore highly entangled in all combinations. It seems reasonable to expect   the nucleus of figure \ref{9} will evolve into the interior of the black hole. In other words after the scrambling time (but long before the Page time) the interior of the  black hole  is the \er \ system that connects the massively entangled near-horizon system of a  black hole.

\subsection{Hawking Radiation}

\begin{figure}[h!]
\begin{center}
\includegraphics[scale=.3]{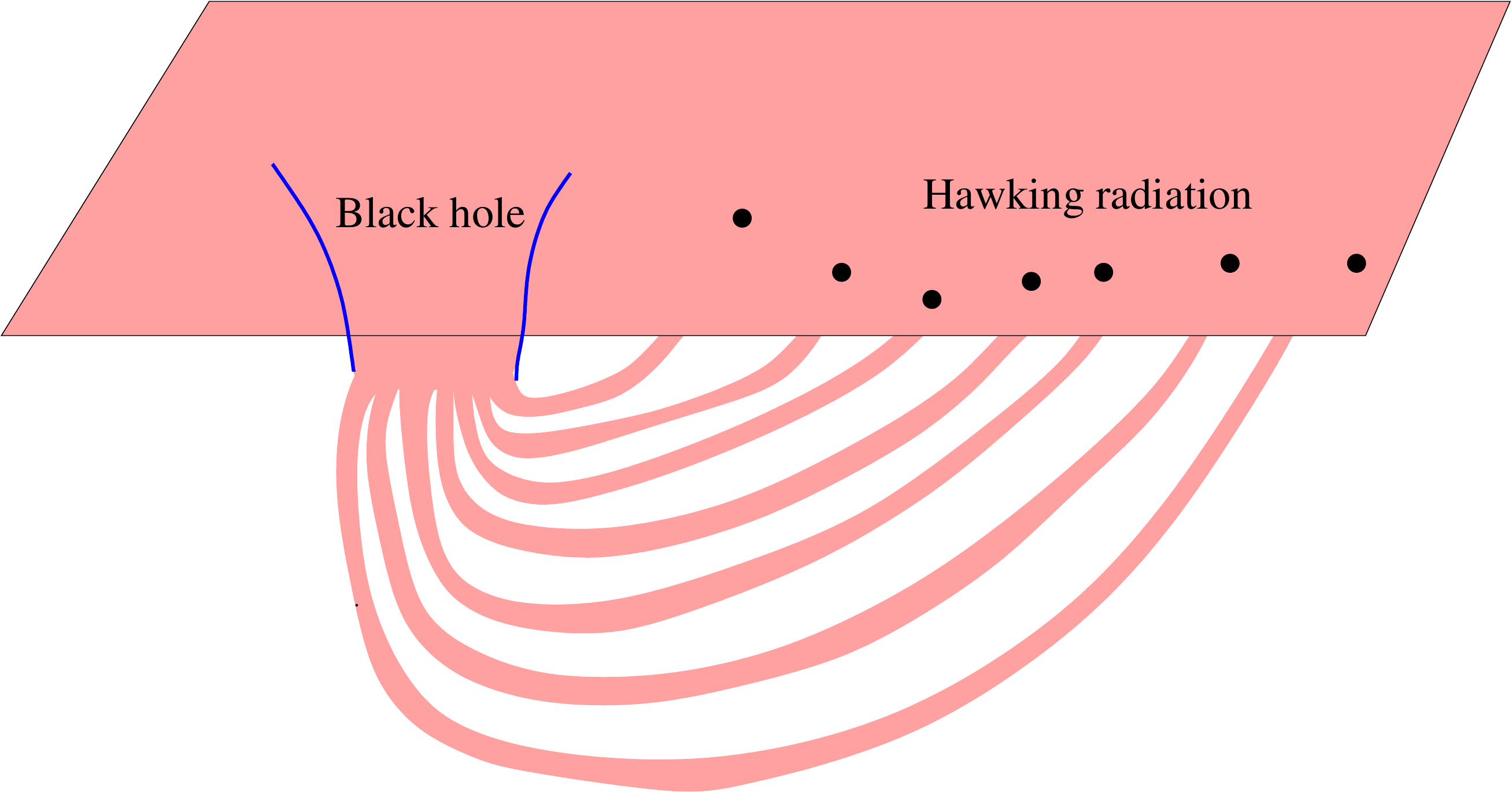}
\caption{ Sketch of the entanglement pattern between the black hole and the Hawking radiation.
We expect that this entanglement leads to the interior geometry of the black hole.   }
\label{BHCloud}
\end{center}
\end{figure}

The Hawking radiation of a black hole is  a scrambled cloud of radiation entangled with the black hole. The obvious configuration of the \er \ would resemble the standard two-black-hole case except that Alice's black hole would be replaced by the Hawking radiation. We can  draw a very impressionistic cartoon of the black hole connected to the radiation by a \er \ with many exits, see figure \ref{BHCloud}.

  \begin{figure}[h!]
\begin{center}
\includegraphics[scale=.3]{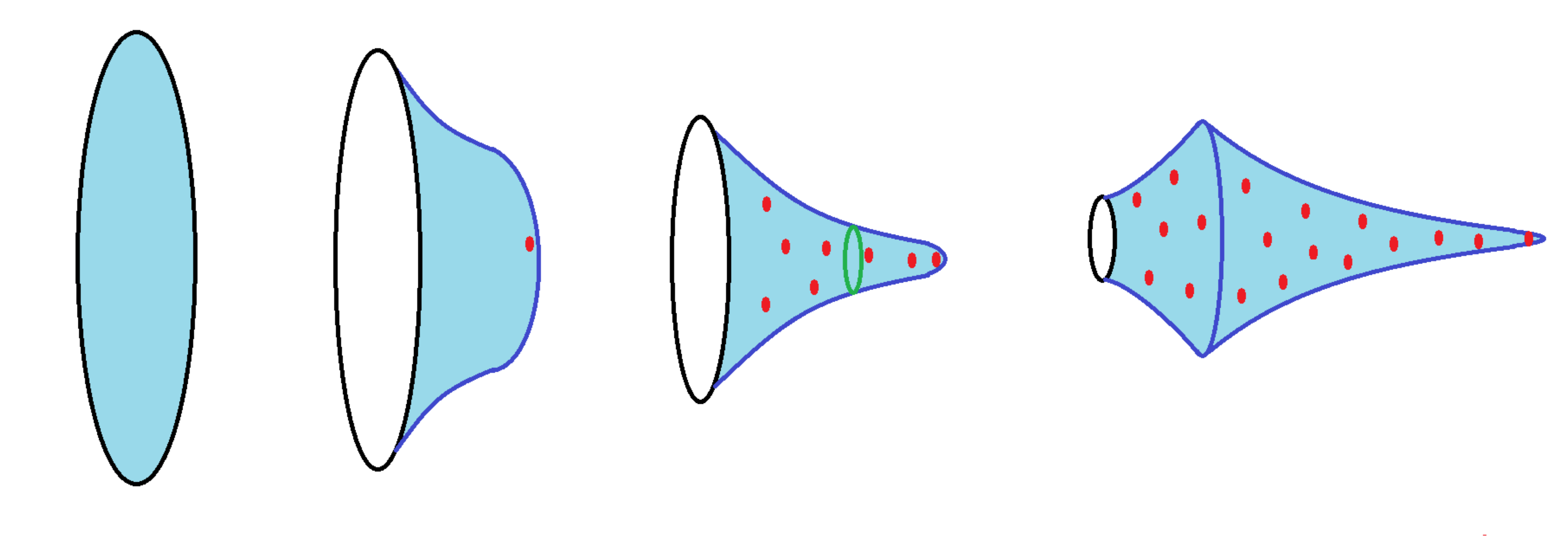}
\caption{ Picture of the evolution of the entanglement through a black hole's life.
% On the left we
%see a black hole that has just formed from a pure state. Here the entanglement is only among different
%black hole degrees of freedom. As the black hole evaporates, we have tiny bridges that provide
%a connection to the Hawking radiation. These are represented as red dots. On the right we see a black hole past its page time, the circle represents the point of maximal entanglement between the subsystem
%to the left and right of the circle.
}
\label{10}
\end{center}
\end{figure}
Another representation is shown in figure \ref{10}.
This figure shows only the geometrical \er \ part of space. On the far left the interior of a young, one-sided black hole is shown. The black circle represents the horizon which should be identified with the horizon as seen from the exterior side. In the beginning there is no Hawking radiation.
As we move to the right Hawking quanta are emitted, and since they are entangled with the black hole, they have to be connected to the bridge. The red dots represent the places where the Hawking quanta connect to the main body of the bridge. The earlier quanta are to the right of the later quanta.
The green circles represent slices through the bridge that divide the system into two parts. To the right of the circle the quanta were emitted earlier than to the left. The entanglement entropy across the green circle is approximately the number of quanta to the right. Thus as we slide the green slice to the left the area grows.
The entanglement entropy reaches a maximum at the Page time and then begins to shrink. Thus as more quanta are emitted the horn-like figure begins to shrink. By then the black hole has also shrunk to less than half the initial entropy.

The cartoon is speculative, and is based on the assumption that the bridge has a geometric description. It would follow that the horizon itself is smooth and has no firewall.

\begin{figure}[h!]
\begin{center}
\includegraphics[scale=.4]{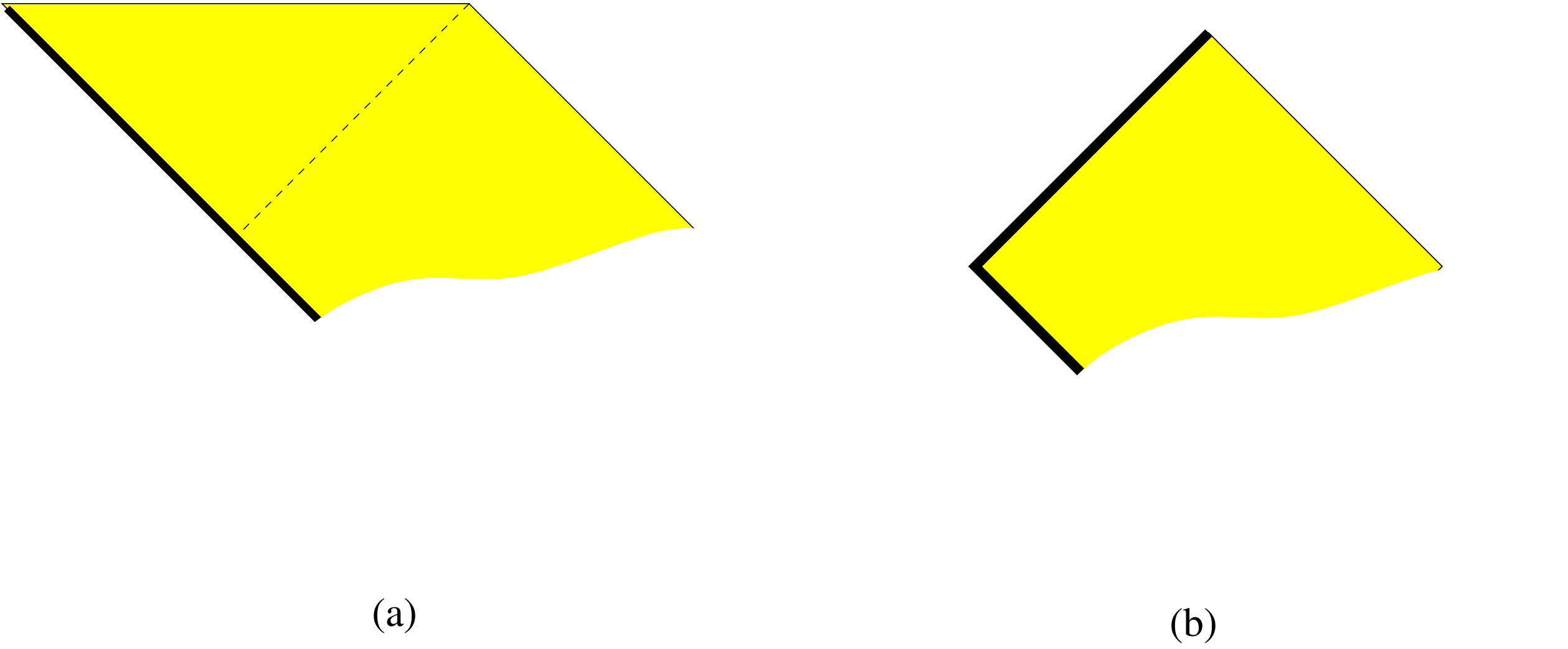}
\caption{(a)  A possible diagram for the bridge connecting an old evaporating black hole to the radiation. The radiation little wormhole mouths would join along the thick black line. When we look at an old black hole we are looking at the upper corner of the Penrose diagram of the original black hole. When we do a time translation (or boost) to focus on the late time region we squeeze the trajectories of the early radiation along the past horizon. (b) The diagram in the firewall scenario, where the smooth geometry stops right behind the horizon. (These diagrams do not take into account the complete evaporation of the black hole).
  }
\label{OldBH}
\end{center}
\end{figure}

 Understanding the \er-bridge system that connects the black hole to the radiation is the key to determining whether the horizons of evaporating black holes are smooth. At the present time we don't know enough to answer the question definitively. See figure \ref{OldBH} for a speculation.

\sc
\section{Implications for the AMPS paradox }

In this section we will examine an AMPS-like paradox \cite{Almheiri:2012rt} in the context of the eternal AdS black hole    where we can use gauge-gravity duality to analyze it. We will construct a situation in which the AMPS argument would lead to a firewall-paradox, and then debug it.

\subsection{Simple and Complex Operators}
Before introducing the setup we will define some terminology. In the various discussions of the AMPS paradox a distinction is  drawn between simple and complex observables \cite{Almheiri:2013hfa}. In the language of qubits the simplicity of an operator represents the number of computational qubits that are involved in its definition. In the black hole radiation the concept of a computational qubit is replaced by the local modes of the radiation field. If we ignore states with more than one quantum in a mode then the localized modes can be replaced by computational qubits. The simple operators in this context are made of a single radiation mode. They are easy to measure or to encode in another system.

By contrast, the operators $R_B$ in \cite{Susskind:2013tg} are extremely complex.  These are the operators that Harlow and Hayden \cite{Harlow:2013tf}  identify as computationally difficult to access.  They are non-locally distributed over the at least half the total number of radiation modes. If the initial entropy of the black hole is $S$ then complex operators involve of order $S$ radiation modes.

In our ADS/CFT-based model we will work in the Schrodinger picture. The  simple units which are easily accessed are the local single-trace operators in the boundary CFT. The most complex operators are very non-local expressions in the gauge theory. They may involve large-scale Wilson loops and even more complicated objects. Experience has shown that the deeper one probes into the interior of AdS, the more complex the probes have to be. An example is the precursor operators in \cite{Polchinski:1999yd}.

\subsection{A Laboratory Example}
\label{labex}

In this section the
 role of Bob's black hole will be played by the  right side of the Penrose diagram \ref{1}: the role of the radiation, by the  left side.  To be concrete we can imagine  a condensed matter system in a fictitious laboratory. The laboratory technician, Alice, will be endowed with extraordinary abilities but only ones which are consistent with quantum mechanics.

Alice has prepared a pair of identical spherical shells, $S_L$ and $S_R$,
 made of a special material that supports a large (but finite) $N$ CFT with a gravity dual.
Note that Alice is not a creature of the bulk. She lives in ordinary space outside the shells; that is, in the laboratory. We will assume that Alice has
 complete control over the shells.   In particular,  by applying a variety of external fields she can manipulate the Hamiltonians of the shells in an arbitrary manner. She can also distill quantum information from a shell, and transport it from one shell to another; all of this  as rapidly and accurately as she likes.  As for the  constraints of special relativity, we may assume that the velocity of propagation on the shells is much slower than the speed of light in the laboratory, so that Alice is unconstrained by any speed limit. She cannot however go backward in time.

Alice has prepared the system at $t=0$ in the maximally entangled  thermofield state so that each side has a black hole. We may, if we like, set up the initial state by assuming a fictitious past consisting of the full Penrose diagram \ref{1} including the lower half.

\begin{figure}[h!]
\begin{center}
\includegraphics[scale=.3]{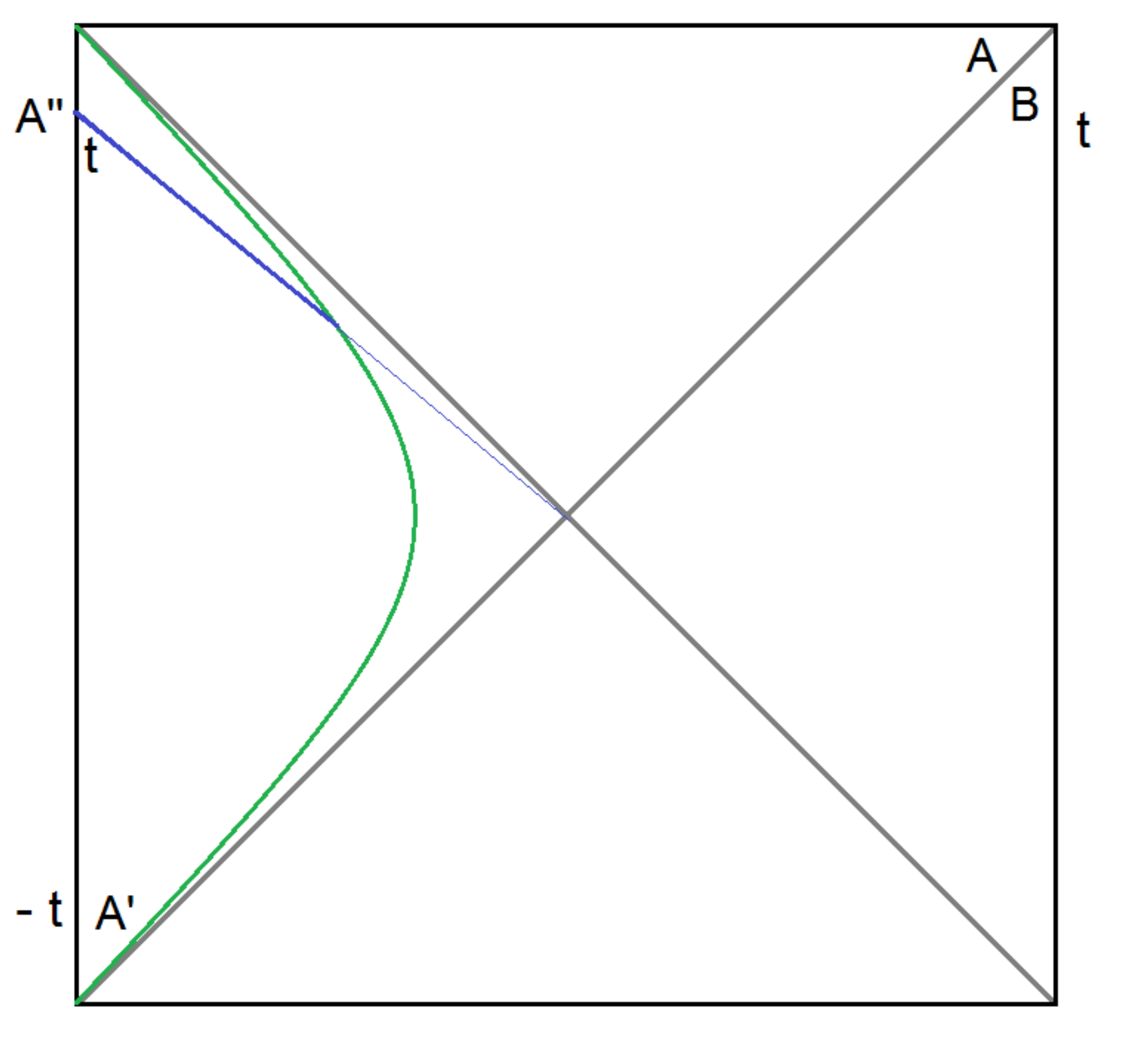}
\caption{  }
\label{16a}
\end{center}
\end{figure}

Bob is a creature of the bulk. He begins life on the right side of the Penrose diagram of figure \ref{16a}. He may or may not cross the horizon into the interior region.

Following AMPS we also consider a pair of modes, $A$ and $B,$ situated on either side of the horizon of Bob's black hole. Both modes correspond to waves propagating toward the upper right, see fig. \ref{16a}.
We can define the time of the $A,B$ pair as the time  where $B$ arrives at the right boundary.

The $A,B$ modes can be considered to be two entangled qubits. They correspond to modes with a wavelength
larger than the Planck scale and an energy comparable to the temperature, so that their entanglement is
close to maximal.
 %Having left the stretched horizon shortly before, at time $t,$ the mode $B$ is in the zone with a wavelength larger than the Planck or string scale.
Following AMPS \cite{Almheiri:2012rt} we may follow $B$ out to the boundary (asymptotic infinity in the AMPS case) and identify $B$ as a simple operator in the right-side CFT.

In the thermofield state, $B$ is maximally entangled with a mode on the left side of the Penrose diagram, at a point  obtained by rotating the diagram by 180-degrees about the origin. That brings us to the operator $A'$ at time $-t.$ Evidently $B$ is entangled with both $A$ and $A',$ but there is no contradiction since $A$ and $A'$ are not independent. In fact  the Penrose diagram suggests that we
 obtain $A$ from bulk evolution from $A'$. Note that this evolution could depend on
 signals that come from the right side. Namely, an object falling from the right could flip the
 qubit, for example. We then write
\be
 A' \longmapsto A
\label{a=aprime}
\ee
indicating that $A'$ determines $A$ through bulk evolution.
%We have assumed that we can use the bulk equations to go from $A'$ to $A,$ and that the field is described by free field equations.

One may question the use of low energy bulk equations, particularly when $t$ is very late.  Replacing $A'$ by an operator on the initial condition surface $t=0$ would involve exponentially small wavelengths. We will come back to this point. For the moment we can Lorentz boost the diagram so that $B$ and $A'$ are transformed to $t=0.$ In this configuration no large energies are encountered in propagating from $A$ to $A'.$

Keeping in mind that $A$ is a qubit, and therefore has three components (the Pauli operators) we write the commutation relations
\be
[A_i, A_j] = i \epsilon_{ijk} A_k \neq 0
\label{aa=a}
\ee
Using \nref{a=aprime} allows us to write
\be
[A'_i, A_j] \neq 0.
\ee

Let us return to the question of replacing $A'$ by an operator on the initial condition surface $t=0.$ In quantum field theory this procedure would involve solving the equations of motion and expressing $A'$ in terms of operators defined on the $t=0$ surface. By inspection we can see that only operators in the left wedge would be involved. But it is obvious that for large $t$ the result would be a trans-plankian operator of exponentially small wavelength. The bulk procedure would be completely out of control if $t$ is greater than the scrambling time. However, here is where the power of gauge/gravity duality comes into play. The operator $A'$ is a simple local boundary operator in the Hilbert space of the left CFT. In principle the gauge-theory equations of motion allow us to solve for it in terms of operators at $t=0.$

Of course the chaotic nature of the dynamics will insure that $A'$ evolves to a very complex operator if $t$ is greater than the scrambling time, but the evolution is nonetheless well defined.

The gauge-theory equations of motion are far too difficult to actually solve, but the solution must exist. One does not expect it to involve exponentially small wavelengths, but rather a complex  gauge-theory operator. Evidently acting with $A'$ at time $-t$ is equivalent to acting with an extremely scrambled non-local operator at time $0.$ One point to keep in mind is that the evolution of $A'$ does not involve operators in the right CFT.

We can go further and run $A'$ all the ways up to time $t$
from time $-t$ to time $+t$ using the equations of motion and Hamiltonian of the left shell.
The two shells do not interact so running $A'$ forward produces an operator $A''$ at time $t$ which also  lives entirely in the Hilbert space of the left CFT. $A''$ is defined by
\be
A'' = e^{-2iH_L t} A' e^{ 2iH_L t} = U_L A' U^{\dag}_L
\label{app = uau}
\ee
These are Schrodinger picture operators.
$A''$ is the operator at time $t$ that has exactly the same information as $A'$ at time $-t.$ Thus we may write
\be
A'' \longmapsto A ~ .
\label{a=aprimeprime}
\ee
$A''$ is a low energy operator of the same energy as $B.$ It is far from being  a local bulk field, but it acts on the black hole as an operator with energy comparable to the Hawking temperature. It is natural to identify it as a degree of freedom in the stretched horizon of the left black hole. This is quite contrary to the naive expectation that the interior of the right black hole should be built from degrees of freedom in its own stretched horizon.
   We have defined $A''$ as an operator in the boundary theory. We have defined $A$ as an operator in the bulk, which is the bulk evolution of the operator $A'$. The operator $A'$ can be viewed as a
bulk or boundary operator, with the translation being simple in this case. The arrow in
\nref{a=aprimeprime} involves bulk evolution.
It depends on the structure of the bridge or the entangled
state. In other words, if we send a wave from the right that flips the spin in the evolution from
$A'$ to $A$, then in order to be talking about the same $A$, we need to modify $A'$ and thus also
$A''$. 

Another important observation is that $A''$ does not commute with $A.$ We can replace \nref{aa=a}
 schematically by\footnote{   We say ``schematically'' because $A''$ is a boundary operator and
$A$ is  bulk operator. Here we want to emphasize that we cannot distill one without modifying the other.} 
\be
[A''_i, A_j] = i \epsilon_{ijk} A_k.
\label{aapp}
\ee
The operator $A''$  plays the same role as $R_B$ in \cite{Susskind:2013tg}.   Namely, it is an
operator in the radiation subsystem (here the left CFT) that is maximally entangled with the qubit B of the
 right-side black hole.
 Contrary to what is assumed in AMPSS, \nref{aapp} indicates that $R_B$ does not commute with $A.$ The disturbance in $A$ when Alice measures or distills $A''$ will play a central role in understanding
  the AMPS paradox.
In AMPS there were two distant systems,  the black hole and the radiation that was entangled with it.
The qubit  $R_B$ ( here $A''$)  was distilled by doing a computation on the radiation,  and then given to
Bob. Here we can do the same. We can distill $A''$, the lab technician can then take it from one
shell to the other and give it to Bob as some excitation introduced from  the UV boundary of his
region.
%In the flat-space case with two black holes on the same sheet, a signal can be sent from the exterior of one black hole to the other. It has to go the long way around and not through the \er.
%
%
%The same is true for the current system. A signal can be sent from near Bob's black hole out to the boundary of his ADS space. From there it can be gathered by Alice and taken to the other side where she may drop it onto the left shell. It than takes some time for it to fall to the near-horizon region of the left black hole. The time for this trip is roughly the ADS radius which, for certain small stable black holes, can be a good deal larger than the transit time across the black hole.  Thus, in the external space the two black holes are distantly separated.

We now have all the ingredients to construct an AMPS-like argument.  At time $t$ The mode $B$ is entangled with a  system, $A'',$ which may be regarded as distant. By the monogamy of entanglement $B$ cannot be entangled with $A.$ Therefore there is a firewall.

The fallacy is clear in this case; namely $A''$ and $A$ are not independent degrees of freedom even though they are very distant in the external space. Indeed, by construction $A''$ determines $A$.
$A$ depends on information from both sides, from both conformal field theories.
 This is the analog of the what in \cite{Susskind:2013tg}  was called $A=R_B.$ Here $R_B = A''$.
  Thus more precisely  what have is  $R_B \longmapsto A $, where the mapping also depends on
  information from the black hole microstates, since an object that falls into the black hole
  can flip the $A$ state.

\begin{figure}[h!]
\begin{center}
\includegraphics[scale=.4]{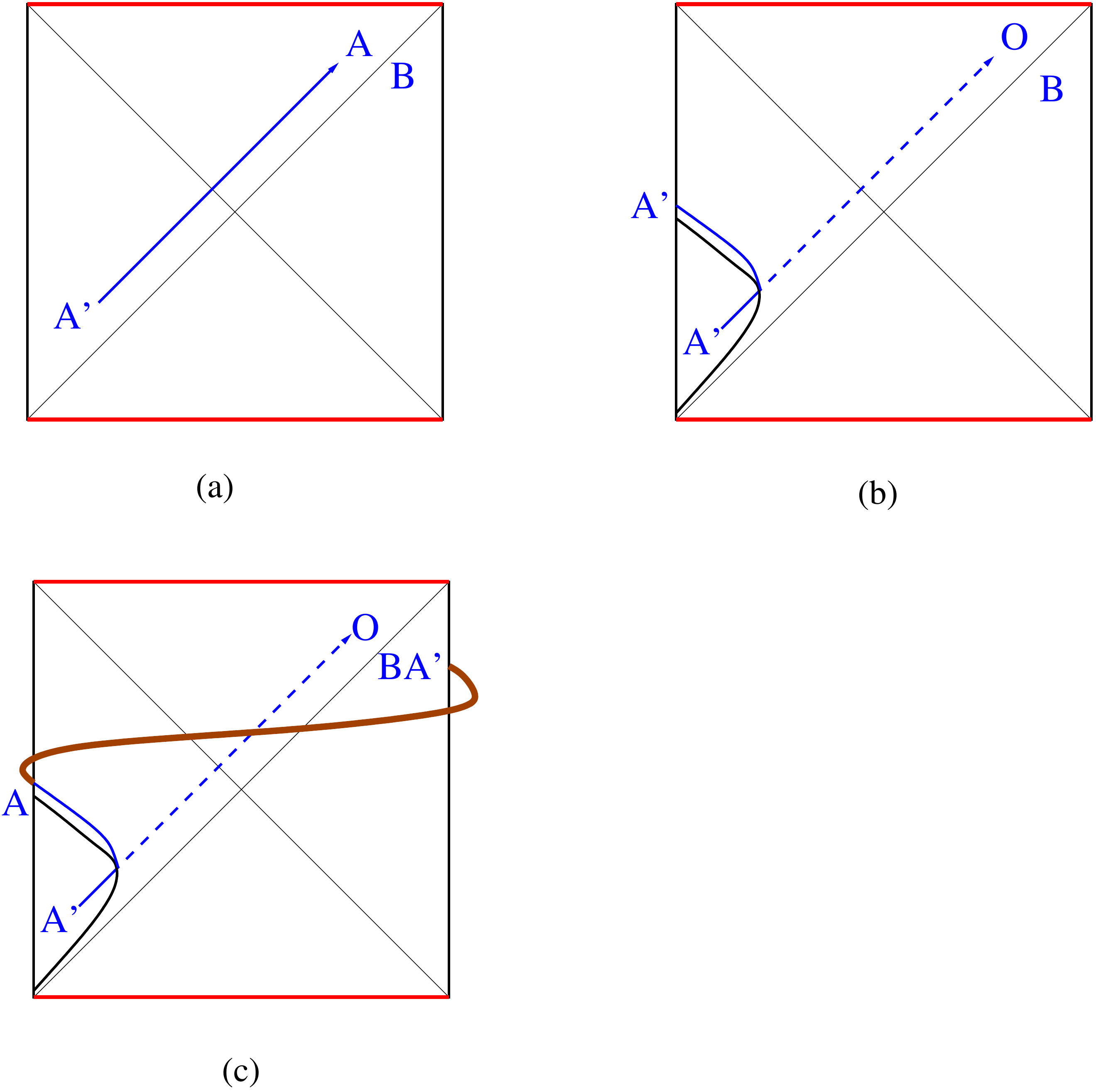}
\caption{ (a) Here we see a mode $B$ which is entangled with $A$. The mode $A$ comes from the
evolution of $A'$. (b) The mode $A'$ is caught by the black bulk object
 and carried to the boundary. This process has
corrupted the mode $A$  and created a particle $O$ on top of the Hartle Hawking state.
(c) A laboratory technician,
 who lives outside the shells,  represented by the thick brown line, takes the qubit $A'$ from one
CFT to the other CFT. There she gives it to Bob. Now Bob sees that $A'$ and $B$ are maximally
entangled. But $B$ is not  entangled with $O$.   }
\label{EternalAMPS}
\end{center}
\end{figure}

Let's follow the AMPS line of reasoning further: first a simplified version, see figure  \ref{EternalAMPS}.
Alice extracts the qubit $A'$
at time $-t$ from the left shell and carries it across to the right shell. She then transfers it into the right shell and gives it to Bob. She does this just in time so that Bob can jump into the black hole
 and test the $AB$ entanglement.
  %for it to  reach the $A,B$ pair.
    We may imagine that Bob carries $A'.$
    Does Bob discover that the $A,B$ system has been corrupted? He does, but the reason is not a firewall. The reason is that $A'$ does not commute with $A.$ When Alice extracted $A'$ she produces a particle in the mode $A.$ That particle is detected by Bob. But the corruption is restricted to a single $A,B$ pair, and does not affect other modes. In any case the corruption  was created by Alice when she extracted $A'.$
This case is simpler because there is no complicated distillation process to be done, the extraction of
$A'$ is fairly straightforward.

It is worth noting that Bob would detect the particle even if he had no knowledge that Alice had distilled $A'.$

Now let us consider what happens if Alice waits until time $t$ and then distills $A''.$
This is a much harder process because $A''$ is a complex operator  which has been scrambled among the stretched horizon degrees of freedom. But by construction it has exactly the same effect as extracting $A'.$ In particular, because of \nref{aapp} it creates a particle in mode $A.$
%
%Note that there is just time to bring $A''$ to Bob's side and throw it into the right black hole.
As before, Bob sees that $B$ is entangled with $A'$ but not with $O$.

The important point is that the quantum detected by Bob was created by Alice when she distilled $A''.$ Contrary to the AMPS assumption, it would not have been there if Alice had not created it in the process.

Note that this example shows  that the problem in AMPS is real. Extracting the qubit entangled
with $B$ from the system that the R black hole was entangled with (which is the L black hole)
 has made it impossible to preserve its entanglement with $A$.

%From the figure  it appears surprising that $A''$ does not commute with $A.$ The stretched horizon of the left black hole %at time $t$ appears to be space-like relative to $A.$ This non-commutativity
%may be related to a similar non-commutativity between naively space-like-separated \dof \ \cite{Lowe:1995ac}.
%In both cases the frames in which the two operators are low energy are related by  large boosts.

This  example demonstrates that entanglement of Bob's black hole with a second system does not imply the existence of a  firewall.

%\subsection{Disturbances from the Right Side}

%The assumption that $A$ involves only operators from the left CFT is not exact. It depended on the assumption of free massless bulk propagation. Interactions of the bulk fields can mix left-moving modes with right-moving modes in the low-energy equations of motion. It is also possible to perturb the right boundary and send in a signal that would affect the propagation from $A'$ to $A.$

%The implication of this observation is that $A''$ contains dependence on the stretched horizon of the right black hole as well as the left. This does not change our conclusions and we will ignore it.

\subsection{Comments on flat space AMPS}

Let us now return to black holes in flat space.
  Consider a black hole that has evaporated past the Page time. Alice has been gathering the Hawking radiation and has more than half of it under her control. Bob has no idea that Alice is out there with her quantum computer. In this version of the argument,
   Alice  takes the radiation and puts it into her quantum computer where she converts it to a black hole. Let's assume for simplicity that the radiation has left Bob's black hole with no off diagonal elements in its density matrix. The far black hole (Alice's) and the near black hole (Bob's) are entangled.  Alice can operate on her black hole
    and create the thermofield state \nref{thermofield state}.
    When Bob jumps in he finds a smooth horizon.

Following AMPS we consider a mode of the Bob's zone which is about to be radiated. The mode is called $B.$
At an earlier time,  Alice has distilled a qubit from her black hole;  namely the qubit that $B$ will be entangled with. In distilling it she has separated it from her black hole and turned it into an ordinary localized qubit.
 Following the notation in \cite{Susskind:2013tg} we call that qubit $R_B.$ With $R_B$ in her possession, Alice flies back to the black hole in time to meet $B$ as it is emitted. As AMPS argued, since $B$ and $R_B$ are entangled, it is not possible for $B$ to be entangled with the mode $A$  (behind the horizon) that it would normally be entangled with. This disruption of the $A,B$ entanglement means that Alice will encounter an unexpected particle as she crosses the horizon.

AMPS then goes on to say that it did not
 matter that Alice carried out the experiment. Even if she had not done so, that high energy quantum would have been there. By that logic, every possible $AB$ pair is corrupted and the black hole horizon becomes a firewall.

We argue AMPS only proved this:

\bigskip

\it  If Alice does the experiment of a given $B$-mode, she will discover that the  corresponding $AB$ entanglement has been corrupted for that one pair.  \rm

\bigskip

 The argument does not prove that any other $AB$ pair is corrupted.
To go further, AMPS must make the assumption that even if Alice did not do the experiment the $AB$ entanglement would be corrupted.

But in our view the particle was created at the other end of the \er \ when Alice distilled  $R_B.$ There is no reason why more than one $AB$ pair should be corrupted by this process. In the eternal black hole
example, we saw this very clearly.

Now Bob will be surprised that he encounters a particle as he goes into the black hole. How did
this particle get there? The answer is through the \er\  that
joins the radiation with the black hole interior. These seem very distant in the external space,
but they can be close via the \er\ bridge.

In this example, we have imagined that Alice has processed the radiation to produce
the thermofield state. Alice can do various things with her quantum computer.
She can arrange to
throw a bomb at Bob.   Alice can
 arrange to get the bomb to Bob in time to intercept Bob as long as Bob in on the inside.
 % If she is uncertain about when Bob will jump in, she can keep resetting the phases and send in a continuous stream of bombs.
 In that case Bob will meet a kind of firewall.

Of course what Alice does is predetermined (in the sense of quantum mechanics) by the initial state of the entire system. But even so, all that can be said, given the exact initial state, is that there is a probability that Bob sees a smooth featureless horizon, and a probability that he gets hit by a firewall, or anything else.

A good question that we can ask is:
  what is the probability for Bob to see any particular thing just behind the horizon, given that a black hole forms from a definite initial state and Alice does not interfere with the radiation? The collapse of a gas cloud in some pure initial state in otherwise empty space would be a good example. Is the probability distribution dominated by firewalls or by smooth featureless
  empty space? At the moment both seem consistent. The answer depends on the nature of the \er \ bridge that connects the black hole  system with the Hawking radiation. If it has a smooth geometry, similar to the standard \er \ near the horizon, then firewalls will be improbable. If it is highly singular then smooth horizons will be improbable.
We do not know enough to answer this question. Of course,  effective field theory around a black hole
horizon suggests that the horizon will be smooth. Hopefully,
 we have given enough reasons
for not believing the AMPS argument that there {\it had} to be firewalls.

\sc
\section{Comments on AMPSS and the Construction of the Interior}

In this section we will address  concerns that were expressed by  Almheiri,  Marolf,  Polchinski,  Stanford, and Sully (AMPSS) \cite{Almheiri:2013hfa}.
The concerns have to do with how the interior \dof \ such as $A$ are
 constructed.
First note that our claim is that
the interior is constructed both from the black hole microstates and from
the radiation.
%The interior is realized in a holographic fashion. We think of the black hole
%microstates and the radiation as the ``boundary'' degrees of freedom and the interior
%
The radiation is analogous the left exterior in figure \ref{1}.  The nature of the interior is not precisely known since it depends on the nature
of the \er\ joining the black hole to the radiation.

We will   study related issues in the case of the eternal $AdS$ black hole, where we know
the nature of the bridge. In that case we can pose AMPSS-like questions and resolve them.

\subsection{The AMPSS Experiment}

The AMPSS paradoxes are aimed at showing that the correspondence $R_B \longmapsto A$ is inconsistent.
 %In our opinion the reason for the concern is a too-literal interpretation of the equality of $A$ and $R_B.$ To address the problem we will use the ADS/CFT laboratory model of Section 4 in which $R_B$ is replaced by $A''.$

%The interpretation given by AMPSS is best explained in the Schrodinger picture. AMPSS takes the equality to mean that in any context where the symbol $A$ appears, it may be replaced by a definite function of  Schrodinger-picture gauge-theory operators. (In the flat space case, a definite function of radiation modes.)
%In this very literal interpretation, the function representing a given $A$ never changes. This is not the interpretation suggested by our previous analysis.

AMPSS begins by considering an
 easily accessed simple operator $E$ in the radiation.   For example, $E$ can represent a particular radiation mode in the early half of the Hawking radiation. Or, in the case of two entangled black holes, it could represent a low energy mode in the zone of the left  black hole.  In the laboratory model
 we  represent  $E$  by a local single-trace operator on the left boundary.

AMPSS argue that if $E$ is  one of the qubits that enters into the construction of $A''$ then $ A''$ and $E$ will not commute.
But then, since $A$ is determined by $A''$,   $A'' \mapsto A $,  we can write
\be
[A, E] \neq 0.
\label{com-ae}
\ee

One can therefore expect that the measurement of $E$ will  disturb the $AB$ correlation and create a high energy particle. This, they argue, is not consistent with the fact that the measurement of $E$ took place very far from the black hole. The conclusion is that the particle must have been there even without doing the experiment. Since the argument can be applied to any $A,B$ pair, there must be a firewall.
The main difference between this experiment and the AMPS experiment is that there is no need to distill $ R_B.$

Notice that we had argued in the previous section that the non-zero commutator $[A,A'']$ implied
that a particle had to be present.  In other words, when
we distilled $R_B$ we created a particle.  Thus \nref{com-ae} seems to imply the same.
Below we will explain a crucial difference between measuring $E$ and measuring $A''$.

\subsection{ An AMPSS like experiment for the eternal $AdS$ black hole}

\begin{figure}[h!]
\begin{center}
\includegraphics[scale=.4]{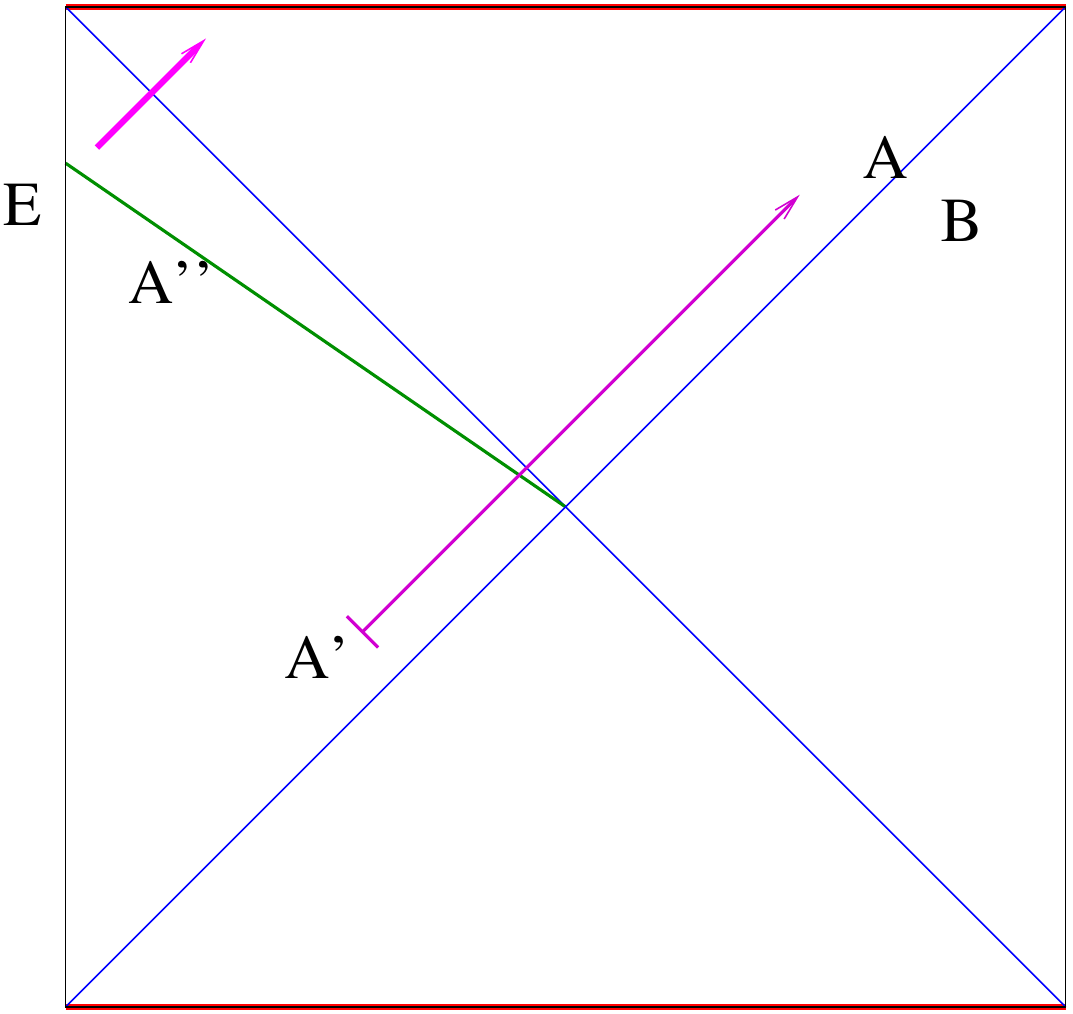}
\caption{ The mode $A'$ evolves to $A$ through bulk evolution. $A''$ is the scrambled operator in
the left CFT that creates the same mode $A'$ but at a later  left boundary time.
At that later time we can also act with a simple operator $E$. According to the bulk picture
this operator creates a simple gravity mode that falls into the singularity and does not disturb $A$.
 }
\label{AMPSSLab}
\end{center}
\end{figure}

Here we consider the same set up as in section \ref{labex}.
The easy mode $E$ will be a low energy gravity mode in the bulk, or a single trace operator
smeared over a region of order the temperature of the black hole.
According to the bulk geometry,
when we measure $E$ at late times on the left,
we create a disturbance in the bulk that moves forwards to the singularity but does not
affect $A$, see figure \ref{AMPSSLab}.
 $E$ is spacelike separated  from $A$ and for this reason they
  commute, at least as far as the bulk is concerned.

From the point of view of the boundary theory we appear to get a different picture.
The operator $A''$ which is the time evolution of $A'$ is quite scrambled, so
 expect that $[E,A'']$ is non-zero. This seems to contradict the bulk picture.

Why do we get two different results? We think that
 the
addition of $E$ changes how we realize $A$ in the boundary theory in terms of
the boundary Hilbert space. However, this change is such that it does not affect the local
physics around $A$. In other words, the physics around $A$ is described in terms of a local
Hilbert space  $H_A$. This Hilbert space describes the presence or absence of various modes
in the neighborhood of the bulk region where $A$ is located. This Hilbert space is embedded
in the boundary Hilbert space in a way that depends on the bridge, on the shape of the spacetime
connecting it to the boundary.
As we act with $E$ we are changing the shape of this bridge and the way in which $H_A$ is embedded.
However, this change is such that we can still extract the underlying state in $H_A$ that we had
in the bulk.
%This will not be the case for all possible boundary operators. For example, if
%we acted with $A''$ we would change the state in $H_A$. But it should be possible when we
%act with the easy operators $E$.
%Bulk locality seems to be  the statement that the $H_A$ space can be moved around in the
%big boundary Hilbert space, but in a way that does not affect the state within $H_A$. We can think of
%this as a sort of ``quantum equivalence principle''.
In the next subsections we hope to explain this  in more detail by giving simple examples.

%However, the bulk is telling us that, as far as the bulk is concerned, the commutator is effectively zero.
%We think that the main point is that the action of  $E$  has not destroyed the information
%contained in $A''$, as we explain in more detail in the next section.

%The fact that this bulk commutator is zero depends on the nature of bridge. Since we do
%not know the nature of the bridge for the case of the black hole entangled with radiation, we
%cannot say whether this bulk commutator is zero or not.

Note that this issue is similar to the problem of bulk locality in AdS/CFT. We can imagine a glueball
deep inside AdS as analogous to $A$. We can then consider an operator $E$ which is given by a local
operator in the boundary theory, say $T(x)$. Since $A$ has a rather complex expression in terms
of the local degrees of freedom, we might naively expect that $[A,T(x)] \not =0$. In fact, if
$T$ is the stress tensor this commutator is indeed non-zero.
However, the bulk description suggests that this commutator should be effectively  zero.
In this particular case, the answer is simple. The operator $A$, which creates a glueball
in the center of $AdS$ is not gauge invariant. If we were dealing with an ordinary gauge
theory in the interior, then $A$ would be a charged particle, and this charge can be measured
at infinity. Thus, $A$ does not commute with the current $j$ at infinity.
The gauge invariant version of $A$ contains a Wilson line that goes to infinity and
is non-local. This is the reason it does not commute with $j$.
A similar issue arises for the stress tensor and operators that change the energy in the
interior.

In the gravity theory it is not possible to define local gauge invariant operators.
Now, this is usually viewed as a minor nuisance, which should not be important in the limit
that the bulk effective field theory is valid. However, it highlights the fact that a non-zero
commutator in the boundary theory does not translate into an effectively non-zero commutator in the
bulk.
%, see section \ref{difffordiff}.

In order to make sure that we are not doing anything illegal from the bulk point of view, it may be better to think in terms of bulk solutions of the Wheeler de Witt equations.
For each particular entangled state, at a particular value of the boundary time,
 we have a bulk solution to the Wheeler deWitt (WdW) equation that
describes the region that is spacelike separated from the boundary.
Different pure entangled states are different solutions of the WdW equation.
Thus, we have a correspondence between states in the boundary Hilbert spaces and states in
the bulk.

\begin{figure}[h!]
\begin{center}
\includegraphics[scale=.5]{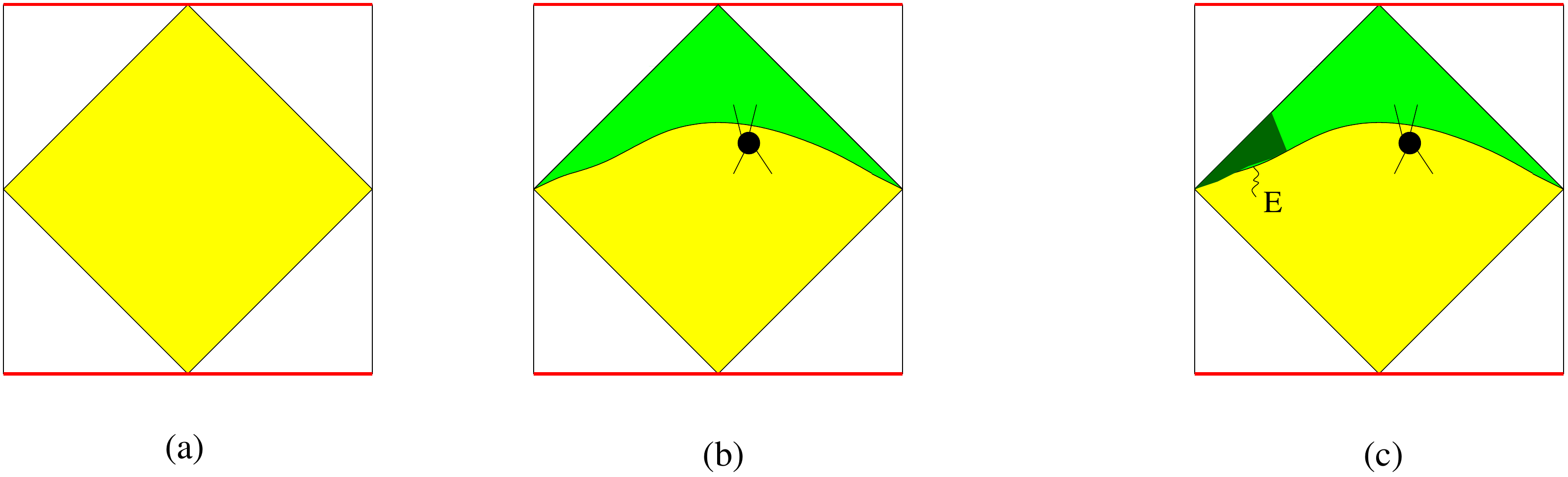}
\caption{  (a) The entangled state at some moment in time corresponds to a solution of the
bulk Wheeler deWitt equation. (b) Here the yellow and green are two different solutions of
the bulk Wheeler deWitt equation that correspond to two different entangled states on the boundary.
Their overlap can be computed both in the boundary and the bulk. In the bulk it can be computed
at any time. If we have a scattering experiment, we can select the yellow wavefunction to have
definite initial states and the green one to have definite final states. Computing their overlap
we can compute the scattering amplitude. (c) Here we added an extra particle near the left
boundary to the yellow wavefunction. We are asking whether it modifies the scattering amplitude.
 In this case we replace the green wavefunction by
  a density matrix of final states which contains any number of particles in the
left corner, indicated by the blob.    }
\label{WheelerdeWit}
\end{center}
\end{figure}

Using this, we can compute the overlap between different states. As usual in the WdW approach, this
overlap can be computed along any spatial slice. In this fashion, we can, in principle, compute
the result of bulk experiments, such as a scattering experiment in the interior. See figure
\ref{WheelerdeWit}(b).

The statement that the commutator is effectively zero in the bulk means that the result of
a scattering experiment in the bulk should not depend on the addition of a particle caused
by the action of $E$, see figure \ref{WheelerdeWit}(c). If the addition of $E$ moves around the
part of the Hilbert space responsible for the region $A$, then this dependence cancels out when
we overlap the initial and final states taking this dependence into account.
% This prescription is following the standard linear rules of quantum
%mechanics both in the bulk and the boundary.

In the AdS problem we took the easy measurement $E$ to be one performed on the left side at late
times, as in figure \ref{AMPSSLab}. Instead,  if we perform $E$ at very early time, in the
bottom left part of the Penrose diagram, we can create a big disturbance in $A$ due to the blueshift factor
\cite{Shenker:2013pqa}.\footnote{ 
Using arguments similar to those of \cite{Shenker:2013pqa} one could argue that the state that
results from adding particles to the thermofield double state will be singular if we evolve it
to the future by a sufficiently long time amount. This arises because evolving the state
backwards in time on the left side is creating a very boosted and energetic particle for the observer
that falls in from the right side. However, by evolving by a Poincare recurrence time to the future
we can have the same effect as evolving backwards in time. Of course, these are extremely long times,
$t_p \sim e^{ e^S } $ so that very strange things can happen. Nevertheless it naively suggests that
 ``half the time'' we would have a firewall \cite{StanfordPrivate}.  }
Whether those $E$'s exist depends, again, on the geometry of the bridge. For example, if the
entangled state is created by black hole pair creation, then the bottom part of the diagram does not
exist. In the case of a black hole entangled with radiation, we do not know.

\subsection{ Information contained in $A$ and error correction}

\begin{figure}[h!]
\begin{center}
\includegraphics[scale=.3]{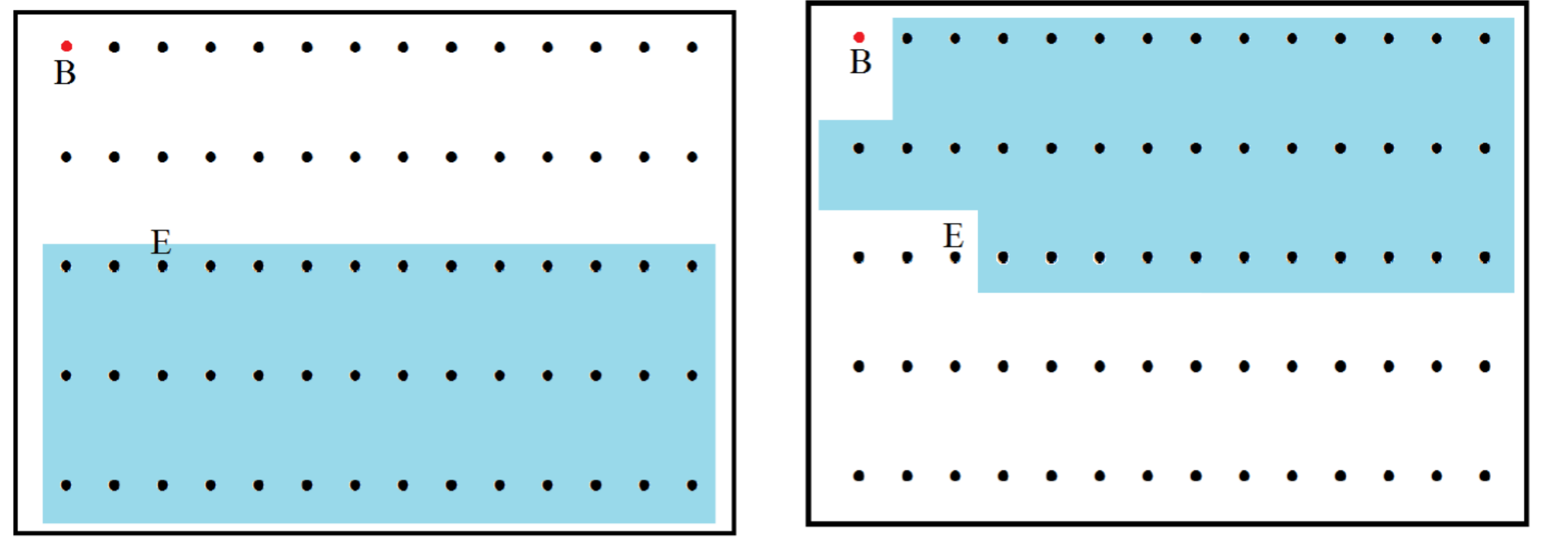}
\caption{  }
\label{16}
\end{center}
\end{figure}

Here we would like to emphasize that the action of the simple operator $E$ does not destroy the
information that is highly delocalized.
As a simple argument
%that an $E$ measurement
% does not destroy the information that we are after
 consider the
following.
Note that $B$ is almost maximally entangled with any subsystem
that contains more than half the total number of degrees of freedom. For example, in figure \ref{16} we show a system of $70$ qubits with the qubit at the upper left corner representing $B.$
Two subsystems, each  with more than half the total number of qubits,  are shown in blue.
Moreover the overlap between the two subsystems can be small.
 In either of these subsystems we can find a qubit that is maximally entangled with $B$, which we are calling $R_B$.
In this way different candidates for $R_B$ can be constructed.

Now assume that $E$ was contained in the construction of $R_B.$ One can always choose another $R_B'$ that does not involve $E.$ $R_B'$ is also almost maximally entangled with $B$ and contains the same quantum information as $R_B$ to a very high approximation.
But
\be
[R_B', E] =0.
\ee
Thus if we identify $R_B'$ with $A$, the measurement of $E$ does not disturb the $AB$ system.

%So we see that there is a large amount of redundancy in the construction of $R_B$.
%To understand the redundancy better,  consider the Hilbert space $V$ of the radiation \r.
% If the number of qubits in \r \ is bigger than the number in the
%  black hole (after the Page time) then the black hole is entangled with a subspace of $V$, call it   $v.$  The dimensionality of $v$ is the same as that of the black hole, and smaller than $V.$

%For each $B$, we have different operators $R_B$ which act on qubits that are maximally
%entangled with $B$.
% The different $R_B$ operators (associated with the same $B$)   are not the same as one another.
% These various $R_B$ operators (all associated to the same $B$)
%  have two important features. First of all when that act on a state in $v$ the result remains in $v.$ Secondly, within $v$ they all act the same way.

% The AMPSS problem is that although the $R_B$  preserve the subspace $v$, the simple operators $E$ do not. Thus when $E$ acts
% the resulting state will be outside the subspace $v,$ and when the various $R_B'$ act on it, they do not behave the same way. However the subset  which commute with $E,$  all do act the same.

 We suggest that the correct prescription is to always employ
 %an $R_B$ which does not lead out of $v.$ There are many of them and they all behave the same way on $v.$ In particular if a measurement of some set of $E$ has occurred we can think of it as causing an error. The error may be corrected by choosing an $R_B$ operator that commutes with
 an $R_B'$ that commutes with $E$.

% With this prescription for how to represent the $A's$ there does not seem to be any observed non-linearity in the infalling frame. Consider for example a correlation function
% \be
% \la A_i, A_j,...,  B_p ...\ra
% \ee
% of operators in the zone and interior. The $B$ do not lead out of the subspace $v,$ and with our prescription neither do the $A.$ Thus there is no dependence of such correlation functions on whether a measurement of $E$
% took place in the radiation degrees of freedom.

As another argument that the information contained in $A$ is not destroyed by the easy measurement $E$,
we can make an analogy with quantum error correction. The connection with error correction was also studied in \cite{Verlinde:2012cy}.
% This prescription for choosing $R_B$ might seem a bit arbitrary.
We view $A$ as a message that has been encoded in the radiation. The action of the operator $E$ can
be viewed as an error. These errors act in a simple way in the radiation basis, which we view as
the computational basis.
%However, we can view this
%operation as a quantum error correction procedure. Starting from some state which was acted upon
%by some unknown errors, by the action of some $E$s,
%one can correct these errors and restore the quantum state of the coded message.
%For this it is important that the errors are not arbitrary operators but that they belong so a
%suitable subset. For example, we can restrict to errors which flip a few qubits at a time in
%the computational basis, which is here   the radiation basis.
% Let us return to the flat space black hole entangled with its radiation.
%We can view  the measurement of $E$ as an error in a
%quantum computation.
%We can view the radiation as the easily accessible basis of qubits. The operator $E$ acts in
%a local way on this basis. It acts on a few qubits at a time. We then view $A$ as some quantum
%information that is present in the state of radiation. This information is present in a
%very scrambled fashion in the state of radiation. Notice that the number of qubits that we need
%to describe the interior is less than half the total number of qubits of radiation.
Because the radiation system is in a random state one expects the encoding of $A$ to be fault-tolerant \cite{Hayden:2007cs}. This means that is it possible to correct the errors.
One can correct errors which act on less than $n/4$ qubits \cite{HaydenPrivate}. For the case that we measure all of the
outside radiation modes, we do not know what happens to the interior, its fate might depend on
the nature of the entanglement with the measuring apparatus.

 By including an ancillary system, which starts
out in a known quantum state, one can represent the whole error correcting procedure as
a unitary operation \cite{PreskillBook}.
 It is thus possible to represent the operators $A$ which act on the
  quantum corrected state as $A_{corrected} = {\cal U}^{-1} A {\cal U}$, where
  ${\cal U}$ is the unitary operator that corrects the errors. Notice that ${\cal U}$
  depends on the whole set of errors that we wish to correct for, but it is independent of the
  particular error that has actually occurred. Here ${\cal U}$ acts on the radiation Hilbert space
  plus the ancillary system.
  In our case, we are not sure exactly what plays the role of the ancillary
system\footnote{ It could be the part of the vacuum state outside the black hole which is not populated by
Hawking radiation.}.

%In AMPSS, they also consider the dependence on the initial state that formed the black hole.
%Notice that according to  \cite{Hayden:2007cs} the information about the initial state has already
%come out by the Page time. Thus, from the point of view of this section the initial state

In summary, our main point is that the interior qubits are very non-locally encoded in the
combined radiation plus black hole microstates system.
 Simple local operations do not destroy the quantum
information contained in them. We can view simple measurements on the radiation as
errors that can be corrected.
In this  picture,
the construction of the interior from the radiation has
to incorporate such an error correction procedure. We do not claim to know what this map
 is explicitly, or how to construct it in practice. All we are saying is that this {\it could}
 be the way the black hole interior is constructed.
 Note that this can isolate the interior from simple
measurements on the radiation, but it does not isolate the interior from very {\it complex} measurements
or operations on the radiation. We still think that if one performs the AMPS operation of
distilling the $A''$ qubit, then we will indeed change the interior.

In summary, the difference between $A''$ and $E$ is that the action of $A''$ can destroy the quantum
information contained in $A$, while $E$ does not destroy it. When we act with $E$,
the information is still in the system and it can be recovered.

\subsection{Linearity}

Another interesting issue discussed in AMPSS has to do with the apparent violation of the linearity of quantum mechanics.
The issue was also raised in \cite{Susskind:2013tg}. It can be stated as follows:

In quantum mechanics observables are represented by linear operators and those operators do not depend on the state of the system. For example the identification $p = -i\hbar\frac{\partial}{\partial x}$ does not depend on what wave function $p$ acts on. If the operator did depend on the state we would say that it was a non-linear operator.

Now consider an observable geometric feature, say a bump in space. The logic of quantum mechanics requires that the feature is associated with
  linear operator; for example something built out of operators representing the  curvature tensor.

On the other hand, entanglement is not a linear feature. We can build an entangled state out of
a linear combination of unentangled states. However, if we have a {\it specific}  entangled state, then
we can consider a projection operator into that specific entangled state. For example, in the case
of two qubits we can project on to the total angular momentum zero state. Or we can project onto any
of the spin one states.

In that light consider $ER = EPR.$ It is important to recall that each particular entangled
state will give a particular \er , see section \ref{difffordiff}. Thus, a given bridge, which
is a specific geometric feature has an associated linear projection operator. This projection
operator acts on the product of the two Hilbert spaces to give a specific entangled state and
a correspondingly specific bridge geometry.

% One might expect that there is an operator $K$ which takes the value $1$ if there is a bridge between two systems, and $0$ if there is not. But we have claimed that the criterion for a bridge is that the two systems be entangled and that any entanglement represents the existence of a bridge, no matter how singular. Thus an \er \ is a geometric feature that is identified by the existence of entanglement. Entanglement is not a linear property. For the case of two qubits there is a complete set of states that has zero entanglement, but it does not follow that all states are un-entangled. It follows that the existence of \er \ does not define a linear subspace of states.

%This seems strange, but it is not. What {\it is} a linear feature is the existence of a
%particular kind of bridge. As we discussed in section \nref{difffordiff} each particular entangled
%state corresponds to a particular bridge, or a particular geometric feature.

%In AMPSS it was argued that, for an old black hole,
% one cannot define the interior creation  with negative
%``energy'' from the microstates of a single black hole.
% We agree.
% In the case of the eternal black hole, it seems clear that one needs the microstates
% of both black holes to define it. For a generic old black hole it is then natural to
 %expect that one could construct it using both the black hole microstates and the radiation.
% Its precise expression depends on the nature of the bridge, but, as we discuss below this
 %seems consistent with quantum mechanics.

\subsection{A  Comment on phases }

Let us now make some remarks on relative phases in the construction of the
Minkowski vacuum from Rindler modes.
As is well known, the Minkowski vacuum has the following expression in terms of the
Rindler vacua
\be \label{mink}
|0 \rangle_M = \exp \left\{   \int d \omega  e^{ -\beta \omega/2} b^\dagger_{L,\omega} b^\dagger_{R,\omega}
 \right\} | 0 \rangle_R
\ee
We can consider the action of a unitary transformation that adds some phases in this relation
\be \label{phasedef}
U_\theta | 0 \rangle_M =  \exp \left\{  \int d \omega e^{ -\beta \omega/2}
e^{ i \theta(\omega) }  b^\dagger_{L,\omega} b^\dagger_{R,\omega}
 \right\} | 0 \rangle_R
\ee
For any non-trivial $\theta$ these are   states that  are boost invariant and differ from the
Minkowski vacuum. All such states carry infinite Minkowski energy\footnote{
Note that $[K,[K,E_M]] = E_M$ where $K$ is the boost and $E_M$ is the Minkowski energy. Thus
if the state is boost invariant, we either have $E_M =0$ or infinity. }.
Thus, for such states we cannot neglect the gravitational back reaction.

One could imagine relaxing the perfect boost invariance so that the Minkowski energy is not
infinite. In other words, we can act with a unitary operator on the Left side in such a way
 that we create a state with finite Minkowski energy. We can ask what this operation corresponds
 to in the $AdS$ example.
 % JM
 For simple\footnote{We are not making any statement about   ``generic'' unitary
 operators.}  unitary operators
 such states are expected to add additional particles on top of the
 Hartle-Hawking vacuum, as discussed in section \ref{difffordiff}.
% JM

\begin{figure}[h!]
\begin{center}
\includegraphics[scale=.4]{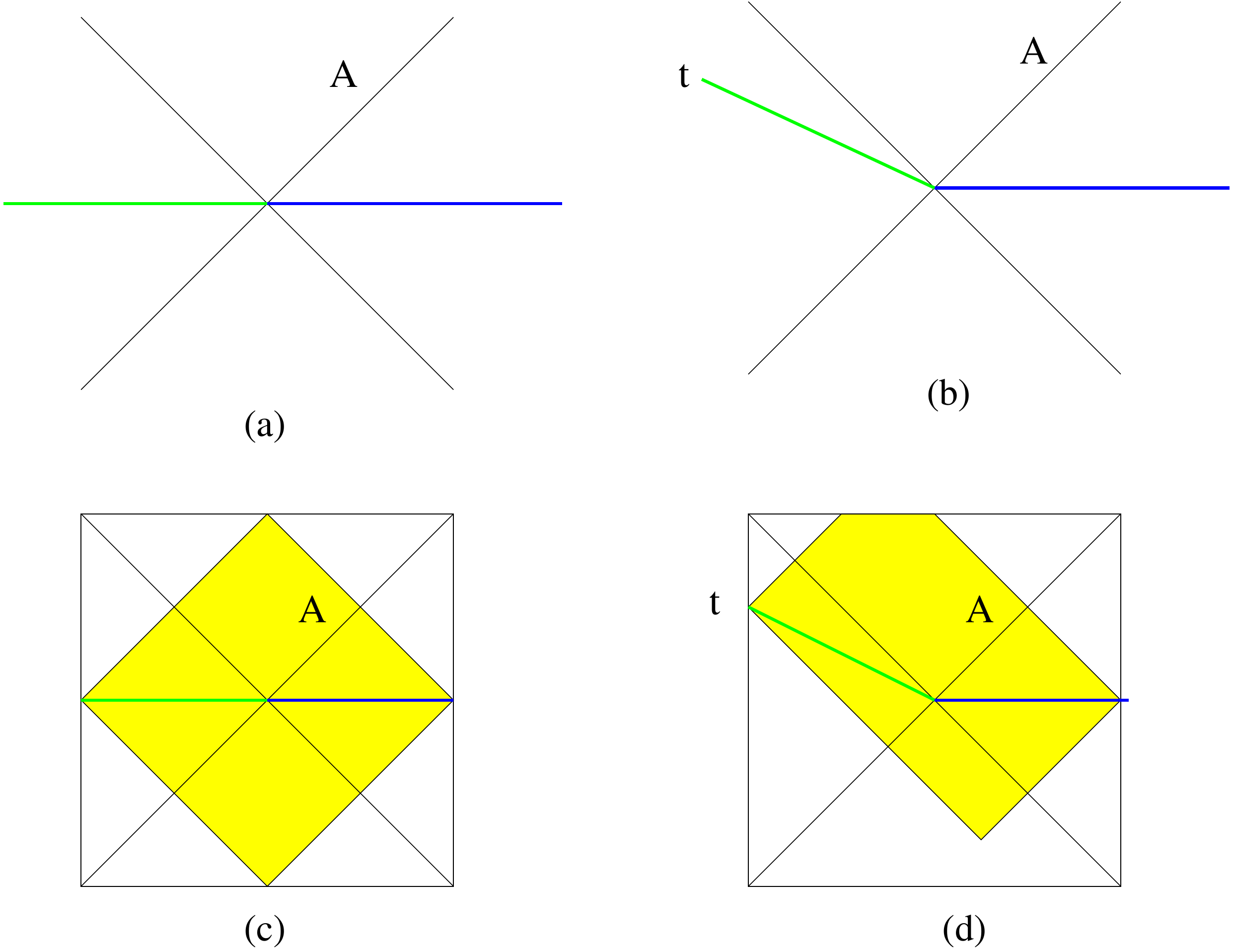}
\caption{ (a) Minkowski vacuum in terms of the Rindler  modes along
the left and right Rindler spatial slices. Along these slices, the state \nref{mink} is regular and
but \nref{kinkstate} is singular. (b) We boosted the left slice. Now the state \nref{kinkstate} is
regular. In (c) we consider the usual thermofield state \nref{thermofield state}. In (d) we consider the result of
adding a left time translation to the thermofield state \nref{t-dependence}.
We get a regular state.   }
\label{Kink}
\end{center}
\end{figure}

Let us study the gravitational back reaction in one very particular case.
Imagine that what we do is to add the phase $\theta(\omega) = - \omega t $ so that we consider
the particular state
\be \label{kinkstate}
U_\theta | 0 \rangle_M =  \exp \left\{  \int d \omega e^{ -\beta \omega/2}
e^{-  i \omega t  }  b^\dagger_{L,\omega} b^\dagger_{R,\omega}
 \right\} | 0 \rangle_R
\ee
This is a state with infinite Minkowski energy.

However, this state, \nref{kinkstate},
 can also be viewed as the expression for the Minkowski vacuum but quantized
along a different spatial slice, a slice with a kink as in figure \ref{Kink}(b).
Thus the state \nref{kinkstate} is very singular if we view it as quantized along the
slice in figure \ref{Kink}(a), but non-singular along the slice \ref{Kink}(b).
Now, if we take this second point of view, we see that an operator in the future, call it $A$,
 has a different
expression in terms of these $a$ and $a^\dagger$ operators  than the one that we have
in the usual case, when we quantize along the usual line, \ref{Kink}(a).
So, the same spacetime region has different expressions in terms of Schrodinger picture
operators if we choose one slice vs. the other.  Now, here we would not call this a ``state dependence''
because we are simply talking about the same abstract state but with different choices for the
parametrization of the operators.

%Now let us go back to the two copies of the conformal field theory.
%In that case we can also consider the family of state that is somewhat analogous to
%\nref{kinkstate}.

Now, let us return to the case of the two sided AdS black hole.
When we do time evolution on one of the
sides we also introduce relative phases between the states of the left and the right CFT
 \nref{t-dependence}.
% In the boundary theory we get  simple phases.
But we do not get a singular state in the bulk. The phases
 reflect in the bulk by constructing a different bridge, a different region of
the black hole spacetime, see figure \ref{DifferentBridges}(c). Interestingly, this different region is such that the state given by
these phases is no longer singular. In some sense,  the interior automatically adjusts so that
we are effectively quantizing along the surface with the kink so as to get a non-singular state.
Now we can consider an $A$ operator which is in the interior and that is common to both
regions. The construction of $A$ in terms of Schrodinger picture operators will be different.
But nevertheless $A$ is describing the same spacetime region in the interior.

\sc
\section{Conclusion}

The full Kruskal geometry figure \ref{2} with its two asymptotic regions has a number of interpretations. The most interesting for us is that it represents the geometry of two distant black holes, which have been created in a certain maximally entangled state. Pair creation of near-extremal black holes in an external electric field provides an explicit example of the creation of maximally entangled black holes. % There are many other ways to accomplish this.

We claim that any pair of  entangled black holes will be connected by some kind of  \er.   The \er \ is a manifestation of  entanglement. We further suggest that the connection is more general than just black holes,  although in most cases the bridge may not have a smooth geometric interpretation. The general principle relating \er \ and entanglement we summarized by the relation
$$
ER=EPR.
$$

Black holes created by collapse are usually considered to be one-sided.  But as a normal one-sided black hole evaporates it effectively becomes two-sided at the Page time when it has become maximally entangled with the outgoing radiation. By the ER=EPR principle it  is connected to the outgoing radiation by a complex \er \ with many outlets. The outlets connect the bridge to the black hole, and to the radiation quanta. An observer called Alice can collect the radiation and run it through a quantum computer. The output will be a second black hole maximally entangled with the first, and therefore connected to it by a smooth \er. Here we are ignoring possible limits of quantum computation \cite{Harlow:2013tf}.

% JM
We emphasize that to describe the  interior of a maximally entangled black hole
we need to know how it is entangled. We also  need to consider the system that the black hole is entangled
with. It is not enough to focus on the microstates of the given black hole. The interior is constructed
from the microstates of the black hole as well as the states of the system that purifies the black hole in the full Hilbert space.
  There are many patterns of entanglement that lead to the same
density matrices for the black hole exterior. It is not enough to
know those density matrices; one must know the pattern of entanglement.
 Namely, one needs to know
the particular entangled state.
Even if most entangled states had firewalls, if a black holes is entangled
with a second system in the appropriate way, the firewalls would dissolve and produce a
smooth horizon. This can happen even though the two systems are far away!.
% JM

The argument of AMPS assumes that Alice can distill a particular qubit $R_B$ from the radiation (or from the black hole that Alice has made from the radiation). $R_B$ is the qubit which is maximally entangled with a certain late radiation mode $B.$ After distilling $R_B$ Alice can bring it back to the black hole where she will find that $B$ and $R_B$ are maximally entangled. By the monogamy of entanglement  $B$ cannot be entangled with its partner mode $A,$ on the other side of the horizon. The breaking of the entanglement between $A$ and $B$ implies the existence of a high energy quantum in the mode $A.$ Alice will be hit by that quantum as she passes the horizon. We agree with this conclusion and the
example of the entangled black holes showed this clearly.

There are two possible lessons of the AMPS argument. The first is that in the process of distilling $R_B,$ Alice created the particle in the related mode $A.$ If that's the case there is no need for a firewall. Alice merely created one  high energy particle and found it when she entered the horizon.

The other possibility is that the high energy particle was present whether or not Alice distilled $R_B.$ If so, there must be a high energy quantum in every $A$ mode, i.e., there is a firewall.

AMPS argue that Alice could not possibly have created the particle. For one thing the distillation process took place very far from the black hole, and the particle was detected just beyond the horizon. The second thing is that the particle was moving in the wrong direction to have arrived from outside the black hole. Thus it must have been there all along, at least after the Page time. Applying the  argument to every $A,B$ pair, AMPS concludes that there is a firewall.

We argue that this conclusion is incorrect; the high energy particle in the $A$ mode, we believe, was indeed created by Alice when she distilled $R_B.$ If she distills one such mode she must create at a quantum in the related $A$ mode, but otherwise she  leave the horizon smooth. (Of course it is always possible that a clumsy Alice will accidentally disrupt more than a single $R_B$ and do greater damage to the horizon.)

Moreover, if she does not distill any $R_B$,  no particles will be created and the horizon may be completely smooth. The question is how is it possible that Alice created a particle moving along the horizon, in mode $A,$ from such a large distance?   Our answer is that the particle arrives at $A$ not from outside the black hole, but through the \er. The argument can be made with some confidence in the case where the radiation has been replaced by a second entangled black hole. The most controlled case in which the argument can be illustrated is  the ``laboratory" model of section \ref{labex}. The laboratory model is based on gauge-gravity duality involving two entangled boundary CFT's.

\begin{figure}[h!]
\begin{center}
\includegraphics[scale=.3]{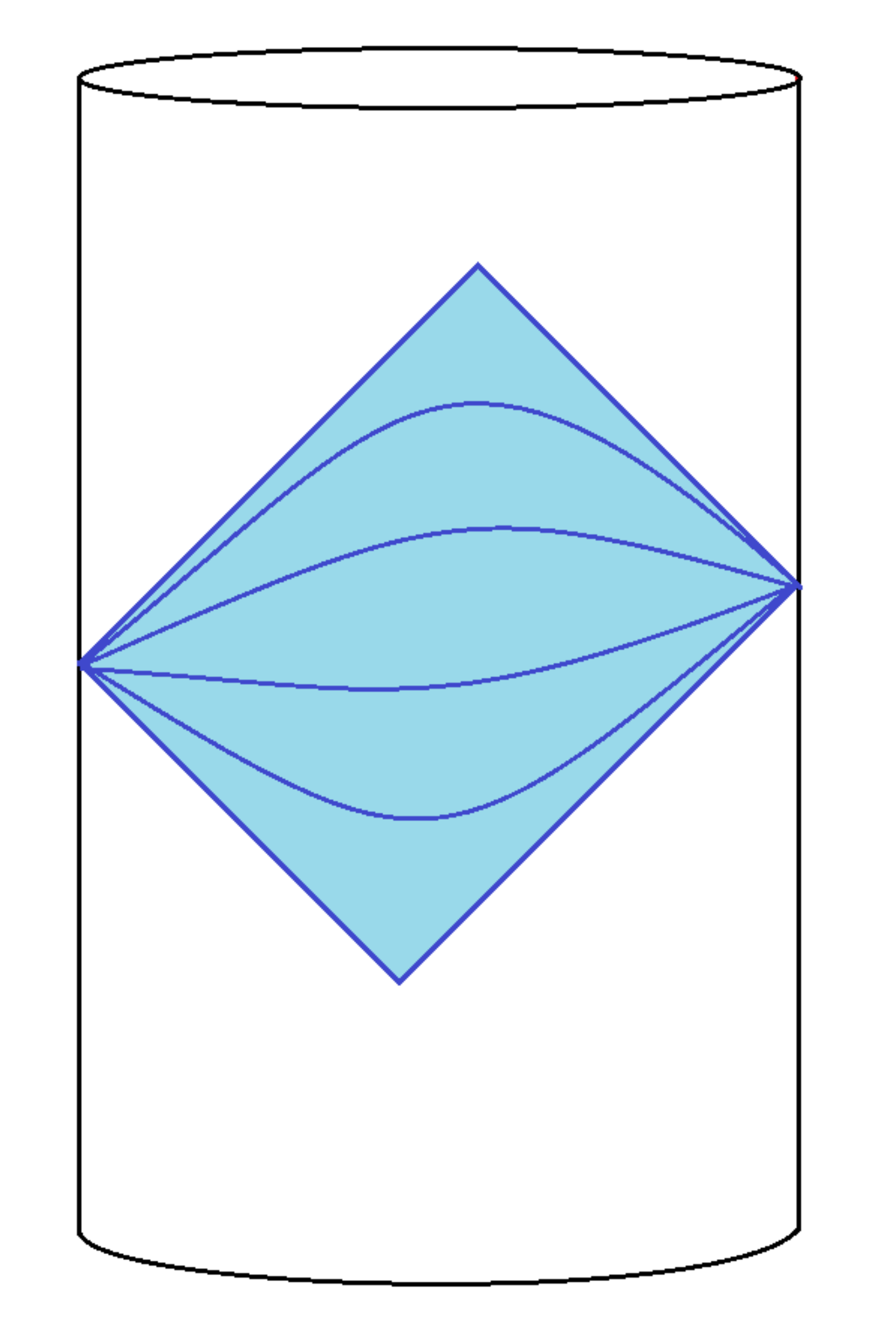}
\caption{ The wavefunction of the boundary theory at some boundary time can be associated to the
whole spacetime region in the bulk that is spacelike separated from that boundary slice. }
\label{final}
\end{center}
\end{figure}

It seems to be a good strategy to formulate behind-the-horizon physics in terms of  Wheeler De Witt wave functions. Consider ordinary ADS/CFT gauge-gravity duality.
The bulk  dual of a Schrodinger picture wave function in the boundary theory is not an instantaneous state in the bulk. It is much better to think of it as the Wheeler De Witt wave function of the whole  space-time region obtained by combining all spacelike surfaces that end at the boundary slice.
 See figure \ref{final}.
The WdW wave-function encodes the bulk in a way that does not depend on slicing.

In the same way,  we should not think of an \er \ as an instantaneous configuration. It is a space-time region of the type shown in figures \ref{final}, \ref{DifferentBridges}, \ref{WheelerdeWit}, \ref{Kink}(d). Different types of \ers \  refer to different types of space-time regions that end on a particular boundary slice. The bridge between two entangled systems
depends on the particular entangled state that we consider, and not  only on the total amount of entanglement. We believe that similar things are true in the case of black holes connected to radiation by an \er.

Regarding the AMPSS paradox, we think that it is important to note that the easy measurements
$E$, that are performed on the radiation, do not destroy the quantum information in the radiation that
is entangled with the near horizon qubits.  The construction of  the mode operator $A$ from radiation qubits is ambiguous and can be done in many ways. AMPSS argue that the action of local operators $E$ does not commute with the operators representing $A.$ From this they conclude that the action of a single $E$ would corrupt all modes and create a firewall. On the other hand we have argued that the large redundancy implicit in a scrambled state allows us to recover from the errors induced by $E.$
We can view the action of $E$ as errors that are acting on a coded message, where the message consists
of the states that the black hole is maximally entangled with. The code consists of all the radiation
qubits which are more numerous than the qubits describing the states entangled with the black holes.
The latter qubits are expected to be highly scrambled in the easily accessed Hilbert space of radiation. The easy measurements, $E$, in AMPSS can be viewed as errors affecting the coded message.
As a matter of principle, these errors can be quantum error  corrected.  Thus, the question of whether
the bulk is destroyed or not is whether gravity is effectively implementing a kind of error
correction procedure automatically or not.

%Another interesting type of redundancy shows up in considering error correction. Construction of the mode operator $A$ from radiation qubits is ambiguous and can be done in many ways. AMPSS argue that the action of local operators $E$ does not commute with the operators representing $A.$ From this they conclude that the action of a single $E$ would corrupt all modes and create a firewall. On the other hand we have argued that the large redundancy implicit in a scrambled state allows us to recover from the errors induced by $E.$ It allows us to construct operators $R_B$ which effectively commute with $E.$

Let us be clear: we do not claim to prove that there are no firewalls, however, we believe that the AMPS conclusion is unwarranted. If it can be shown that the \er \ connecting the black hole to its radiation is smooth near the black hole, then there will be no firewall,  unless Alice's computer performs an extremely difficult operation. At the moment we do not know enough about   \ers \ involving clouds of Hawking radiation to come to a definite conclusion, though general relativity suggests a smooth horizon. However, we claim there is no convincing argument in favor of firewalls.

One thing seems clear: the smoothness of the horizon of an old black hole
 {\it does }  depend on what we do with the radiation.
We can always process the radiation to produce the thermofield double state, which has a smooth
horizon. Or the thermofield double plus extra high energy particles, which do not.

 The AMPS paradox is an extremely subtle one whose resolution, we believe, will have much to teach us about the connection between geometry and entanglement. AMPS pointed out a deep and genuine paradox about the interior of black holes. Their resolution was firewalls, but in our view the solution more likely involves Einstein-Rosen bridges. The interior of an old
 black hole would be an \er \ constructed  from the
micro-degrees of freedom of the black hole  as well  the radiation.

\section*{Acknowledgements}

 We would like to thank N. Arkani Hamed, R. Bousso, D. Harlow, T. Hartman, D. Marolf, J. Polchinski,
  S.  Shenker, D. Stanford and  H. Verlinde   for discussions.
 JM was supported in part by U.S.~Department of Energy grant DE-FG02-90ER40542.
LS was supported in part by NSF under Grant No. PHY11-25915.

\appendix

\section{Black hole pair creation in a magnetic field }

Here we review some formulas from
\cite{Garfinkle:1990eq,Garfinkle:1993xk}.

The background Melvin solution has the form
\begin{eqnarray} \label{melvs}
ds^2 &=& { 1 \over B^2} \left[ \Lambda_b^2 ( dt^2 + dz^2 + d\rho^2 ) + { \rho^2 \over \Lambda_b^2 } d\phi^2 \right]
\\
F &=& { 2 \over B} d \phi d \Lambda_b^{-1} ~,~~~~~~~~~~~~~~~~~~~\Lambda_b \equiv  { 1 + { \rho^2 \over 4} }
\end{eqnarray}
We have pulled out the parameter $B$ so that the scaling  with $B$ is
manifest.
The instanton solution has the form
\begin{eqnarray}
 \label{metrinst}
 ds^2 &=&  { 1 \over A^2} \left[ { \Lambda^2 \over (x-y)^2 } [ - G(y)d\tau^2 - G^{-1}(y) dy^2 + G^{-1}(x) dx^2 ] + { G(x) \over \Lambda^2 (x-y)^2 } d\phi^2 \right]
\\
F& = &  d\phi d \Phi
\\
G(x) &= &  1 -x^2 (1 + q A x)^2 ~,~~~~~~~~~~~~\Phi = { 2 \over \Lambda B } (1 + { 1 \over 2 }q B x )
\\
\Lambda &=& (1 + { 1 \over 2 } q B x )^2 + { B^2 G(x) \over 4 A^2 (x-y)^2 }
\end{eqnarray}
 where $q$, $A$ and $B$ are parameters obeying $qB < 1/4$ and a relation written in eqn. (2.9) of
 \cite{Garfinkle:1993xk}.
 The function $G(x)$ has four zeros, $\zeta_i$, $i=1,\cdots,4$.
 We are interested in the regime where $q B \ll 1$. In this regime we have that $A \sim B$.
 Then the four roots of the equation $G(x)=0$ are
\be
 \zeta_1 = - { 1 \over q A } -1+ \cdots  , ~~~~\zeta_2 = - { 1 \over q A}  +  1+\cdots , ~~~~~~~\zeta_3 =-1+ \cdots , ~~~~~ \zeta_4 = 1+ \cdots
\ee
 We take $ \zeta_2 < y < \zeta_3 < x < \zeta_4 $.

 In the regime where $x$ and $y$ are of order one, we can approximate
  the function  $G(x)$ by
$G(x) =1-x^2$ and similarly for $G(y)$. We then define the new variables
\be
  \rho^2  = { (1-x^2) \over (x-y)^2 }  ~,~~~~~~~~~~ r^2 =  { y^2-1 \over (x-y)^2 }
\ee
We can also approximate $\Lambda \sim 1 + { (1-x^2) \over 4 (x-y)^2} = 1 + { 1 \over 4 }   \rho^2 $.
Then the metric \nref{metrinst} becomes identical to \nref{melvs}, except that
$dt^2 + dz^2 = r^2 d\tau^2 + dr^2$. In other words, we get the metric in polar coordinates.
 These become Rindler coordinates after we analytically continue $\tau$ to Lorentzian time.
The overall scale is set by $B=A$.

This approximate form is correct  as long as we are far from $y \sim \zeta_2 \sim - { 1 \over q A}$.
In that region  cannot longer  approximate $G(y) \sim  1-y^2$. This region corresponds
to $r\sim 1$. This is the region where the black holes are sitting.
More precisely, the metric starts deviating from the Melvin solution \nref{melvs} when
$r-1 \sim  q A $.
In this region, we have
\be \label{largey}
-G(y) = y^2 (1+ q A y)^2 -1  \sim y^2 (1 + q  A y )^2
\ee
where we neglected the 1 since $y$ is large.
Since $x$ continues to be of order one, we can still approximate $G(x) \sim 1-x^2$.
In this  regime we can approximate $\Lambda \sim 1$ and $(x-y) \sim - y $. Then the
metric \nref{metrinst} becomes
\begin{eqnarray}
ds^2 &= & { 1 \over A ^2} \left\{  - {  G(y) \over y^2} d\tau^2 - { 1 \over y^2 G(y)} dy^2 + { 1 \over y^2 } [
{ dx^2 \over (1-x^2)} + (1-x^2) d\phi^2 ] \right\}
\\
ds^2 &= & ( 1 -{ q \over \tilde r})^2 { d\tau^2 \over A^2}  + { d\tilde r^2 \over (1 - { q \over \tilde r})^2 }
+ \tilde r^2 d\Omega_2^2
\\
 && \tilde r = - { 1 \over A y }  ~,~~~~~~~~x=\sin \theta
\end{eqnarray}
Thus the $x$ and $\phi$ directions give a two sphere and the other dimensions
give a black hole with a
Schwarzschild radius equal to $q$.
  This black hole is extremal in this approximation.
When we go near the horizon of the black hole, it is important to remember the 1 in \nref{largey}
this replaces
\be
(1 - {q   \over \tilde r} )^2 \longrightarrow  (1 - { q  \over \tilde r} )^2 - ( q A)^2
\ee
This splits the two zeros by a small amount, giving the black hole a small amount of non-extremality.

The condition that set $B=A$ is the same condition that is setting the temperature of the black hole
to be equal to the Rindler acceleration.
The circular trajectory of the black hole is parametrized by the coordinate $\tau$.

When we view this solution as a tunneling solution, we cut the solution across ordinary Euclidean
time, so that we produce two black holes and the horizon is entangled as expected. Of course, we
produce the wormhole of the near extremal Reissner Nordstrom black hole.
  The Euclidean time circle $\tau$ shrinks to zero at {\it two}
 locations, one
is the horizon of the black hole, which corresponds to $y=\zeta_2$ and the other
corresponds to the origin of space that corresponds to $y = \zeta_3$, or $r=0$.
In this case the ``long'' distance between the two black holes is $d_{\rm long} \sim 1/A$,
while the ``short'' one through the wormhole is $d_{\rm short} \sim - q \log(qA) $. For
$qA \gg 1$ we see that the long distance is much longer than the short one.

The instanton action, as computed by  \cite{Garfinkle:1993xk}, is
\be \label{acti}
e^{ - I } = e^{ - { \pi q \over B} } e^{ { \pi \over 2 }  q^2 }  ~~~~~~{\rm for} ~~q B \ll 1
\ee
For a fixed magnetic field it is most convenient to create the smallest possible
black hole, at least  when $qB \ll 1$.
It turns out that  for the full action, we also maximize the probability for small black holes,
see formula (2.14) in \cite{Garfinkle:1993xk}. In other words, the $q^2$ entropic-like enhancement factor never wins over the first factor. One can isolate the entropic factor by comparing \nref{acti} with
the probability for creating magnetic monopoles \cite{Garfinkle:1993xk}.
That ratio contains the entropy factor
$e^{ S_{BH}} = e^{ \pi q^2 }$, which is similar, but not equal, to \nref{acti} \footnote{The difference is due to Coulomb and Newtonian self interactions of the black hole loop  \cite{Garfinkle:1993xk}.}.
  It is of course, given by the probability of thermally producing
them in Rindler space at the acceleration temperature. Note that, even though we can
``naturally'' produce them in this way, we pay a heavy penalty since the tunneling rate is very small.


\begin{thebibliography}{99}

%\cite{Einstein:1935rr}
\bibitem{Einstein:1935rr}
  A.~Einstein, B.~Podolsky and N.~Rosen,
  ``Can quantum mechanical description of physical reality be considered complete?,''
  Phys.\ Rev.\  {\bf 47}, 777 (1935).
  %%CITATION = PHRVA,47,777;%%
  %892 citations counted in INSPIRE as of 06 May 2013

 %\cite{Einstein:1935tc}
\bibitem{Einstein:1935tc}
  A.~Einstein and N.~Rosen,
  ``The Particle Problem in the General Theory of Relativity,''
  Phys.\ Rev.\  {\bf 48}, 73 (1935).
  %%CITATION = PHRVA,48,73;%%
  %198 citations counted in INSPIRE as of 06 May 2013


%\cite{Fuller:1962zza}
\bibitem{Fuller:1962zza}
  R.~W.~Fuller and J.~A.~Wheeler,
  ``Causality and Multiply Connected Space-Time,''
  Phys.\ Rev.\  {\bf 128}, 919 (1962).
  %%CITATION = PHRVA,128,919;%%
  %41 citations counted in INSPIRE as of 29 May 2013



%\cite{Friedman:1993ty}
\bibitem{Friedman:1993ty}
  J.~L.~Friedman, K.~Schleich and D.~M.~Witt,
  ``Topological censorship,''
  Phys.\ Rev.\ Lett.\  {\bf 71}, 1486 (1993)
  [Erratum-ibid.\  {\bf 75}, 1872 (1995)]
  [gr-qc/9305017].
  %%CITATION = GR-QC/9305017;%%
  %205 citations counted in INSPIRE as of 14 May 2013


 % \cite{Friedman:1993ty,Galloway:1999bp}
 %\cite{Galloway:1999bp}
\bibitem{Galloway:1999bp}
  G.~J.~Galloway, K.~Schleich, D.~M.~Witt and E.~Woolgar,
  ``Topological censorship and higher genus black holes,''
  Phys.\ Rev.\ D {\bf 60}, 104039 (1999)
  [gr-qc/9902061].
  %%CITATION = GR-QC/9902061;%%
  %99 citations counted in INSPIRE as of 14 May 2013


%\cite{Israel:1976ur}
\bibitem{Israel:1976ur}
  W.~Israel,
  ``Thermo field dynamics of black holes,''
  Phys.\ Lett.\ A {\bf 57}, 107 (1976).
  %%CITATION = PHLTA,A57,107;%%
  %258 citations counted in INSPIRE as of 07 May 2013

%\cite{Maldacena:1998bw}
\bibitem{Maldacena:1998bw}
  J.~M.~Maldacena and A.~Strominger,
  ``AdS(3) black holes and a stringy exclusion principle,''
  JHEP {\bf 9812}, 005 (1998)
  [hep-th/9804085].
  %%CITATION = HEP-TH/9804085;%%
  %477 citations counted in INSPIRE as of 03 Jun 2013


%\cite{Horowitz:1998xk}
\bibitem{Horowitz:1998xk}
  G.~T.~Horowitz and D.~Marolf,
  ``A New approach to string cosmology,''
  JHEP {\bf 9807}, 014 (1998)
  [hep-th/9805207].
  %%CITATION = HEP-TH/9805207;%%
  %64 citations counted in INSPIRE as of 03 Jun 2013


  %\cite{Balasubramanian:1998de}
\bibitem{Balasubramanian:1998de}
  V.~Balasubramanian, P.~Kraus, A.~E.~Lawrence and S.~P.~Trivedi,
  ``Holographic probes of anti-de Sitter space-times,''
  Phys.\ Rev.\ D {\bf 59}, 104021 (1999)
  [hep-th/9808017].
  %%CITATION = HEP-TH/9808017;%%
  %263 citations counted in INSPIRE as of 30 May 2013

  %\cite{Maldacena:2001kr}
\bibitem{Maldacena:2001kr}
  J.~M.~Maldacena,
  ``Eternal black holes in anti-de Sitter,''
  JHEP {\bf 0304}, 021 (2003)
  [hep-th/0106112].
  %%CITATION = HEP-TH/0106112;%%
  %289 citations counted in INSPIRE as of 14 May 2013

    %\cite{Almheiri:2012rt}
\bibitem{Almheiri:2012rt}
  A.~Almheiri, D.~Marolf, J.~Polchinski and J.~Sully,
``Black Holes: Complementarity or Firewalls?,''
  arXiv:1207.3123 [hep-th].
  %%CITATION = ARXIV:1207.3123;%%

% \cite{Almheiri:2012rt,Almheiri:2013hfa}
%\cite{Almheiri:2013hfa}
\bibitem{Almheiri:2013hfa}
  A.~Almheiri, D.~Marolf, J.~Polchinski, D.~Stanford and J.~Sully,
``An Apologia for Firewalls,''
  arXiv:1304.6483 [hep-th].
  %%CITATION = ARXIV:1304.6483;%%

%\cite{Braunstein:2009my}
\bibitem{Braunstein:2009my}
  S.~L.~Braunstein,
  ``Black hole entropy as entropy of entanglement, or it's curtains for the equivalence principle'',
  arXiv:0907.1190v1 [quant-ph].
  %%CITATION = ARXIV:0907.1190;%%
  %9 citations counted in INSPIRE as of 03 Jun 2013
  S.~L.~Braunstein, S.~Pirandola and K.~Z.yczkowski,
  ``Entangled black holes as ciphers of hidden information,''
  Phys. Rev. Lett. 110, {\bf 101301} (2013)
  [arXiv:0907.1190 [quant-ph]].
  %%CITATION = ARXIV:0907.1190;%%

%\cite{Mathur:2009hf}
\bibitem{Mathur:2009hf}
  S.~D.~Mathur,
  ``The Information paradox: A Pedagogical introduction,''
  Class.\ Quant.\ Grav.\  {\bf 26}, 224001 (2009)
  [arXiv:0909.1038 [hep-th]].
  %%CITATION = ARXIV:0909.1038;%%
  %65 citations counted in INSPIRE as of 03 Jun 2013



%\cite{Giddings:2012dh}
\bibitem{Giddings:2012dh}
  S.~B.~Giddings and Y.~Shi,
  ``Quantum information transfer and models for black hole mechanics,''
  arXiv:1205.4732 [hep-th].
  %%CITATION = ARXIV:1205.4732;%%
  %10 citations counted in INSPIRE as of 03 Jun 2013


%\cite{Papadodimas:2012aq}
\bibitem{Papadodimas:2012aq}
  K.~Papadodimas and S.~Raju,
  ``An Infalling Observer in AdS/CFT,''
  arXiv:1211.6767 [hep-th].
  %%CITATION = ARXIV:1211.6767;%%
  %13 citations counted in INSPIRE as of 12 Jun 2013

\bibitem{Avery:2013exa}
  S.~G.~Avery and B.~D.~Chowdhury,
  ``Firewalls in AdS/CFT,''
  arXiv:1302.5428 [hep-th].
  %%CITATION = ARXIV:1302.5428;%%
  %3 citations counted in INSPIRE as of 17 Jun 2013.

 %\cite{Cantcheff:2011jk}
%\bibitem{Cantcheff:2011jk}
%  M.~B.~Cantcheff,
  %``Emergent spacetime, and a model for unitary gravitational collapse in AdS,''
%  arXiv:1110.0867 [hep-th].
  %%CITATION = ARXIV:1110.0867;%%
  %8 citations counted in INSPIRE as of 17 Jun 2013

    %\cite{Harlow:2013tf}
\bibitem{Harlow:2013tf}
  D.~Harlow and P.~Hayden,
 ``Quantum Computation vs. Firewalls,''
  arXiv:1301.4504 [hep-th].
  %%CITATION = ARXIV:1301.4504;%%
  %4 citations counted in INSPIRE as of 03 Mar 2013



%\cite{Garfinkle:1990eq}
\bibitem{Garfinkle:1990eq}
  D.~Garfinkle and A.~Strominger,
  ``Semiclassical Wheeler wormhole production,''
  Phys.\ Lett.\ B {\bf 256}, 146 (1991).
  %%CITATION = PHLTA,B256,146;%%
  %83 citations counted in INSPIRE as of 07 May 2013


%\cite{Garfinkle:1993xk}
\bibitem{Garfinkle:1993xk}
  D.~Garfinkle, S.~B.~Giddings and A.~Strominger,
  ``Entropy in black hole pair production,''
  Phys.\ Rev.\ D {\bf 49}, 958 (1994)
  [gr-qc/9306023].
  %%CITATION = GR-QC/9306023;%%
  %80 citations counted in INSPIRE as of 07 May 2013

%\cite{Heemskerk:2012mn}
\bibitem{Heemskerk:2012mn}
  I.~Heemskerk, D.~Marolf, J.~Polchinski and J.~Sully,
  ``Bulk and Transhorizon Measurements in AdS/CFT,''
  JHEP {\bf 1210}, 165 (2012)
  [arXiv:1201.3664 [hep-th]].
  %%CITATION = ARXIV:1201.3664;%%
  %18 citations counted in INSPIRE as of 12 Jun 2013


%\cite{Bak:2007qw}
\bibitem{Bak:2007qw}
  D.~Bak, M.~Gutperle and A.~Karch,
  ``Time dependent black holes and thermal equilibration,''
  JHEP {\bf 0712}, 034 (2007)
  [arXiv:0708.3691 [hep-th]].
  %%CITATION = ARXIV:0708.3691;%%
  %13 citations counted in INSPIRE as of 21 May 2013
 %\cite{Bak:2007jm}
   D.~Bak, M.~Gutperle and S.~Hirano,
  ``Three dimensional Janus and time-dependent black holes,''
  JHEP {\bf 0702}, 068 (2007)
  [hep-th/0701108].

\bibitem{Ryu:2006bv}
  S.~Ryu and T.~Takayanagi,
  ``Holographic derivation of entanglement entropy from AdS/CFT,''
  Phys.\ Rev.\ Lett.\  {\bf 96}, 181602 (2006)
  [hep-th/0603001].
  %%CITATION = HEP-TH/0603001;%%
  %285 citations counted in INSPIRE as of 07 May 2013


%\cite{VanRaamsdonk:2010pw}
\bibitem{VanRaamsdonk:2010pw}
  M.~Van Raamsdonk,
  ``Building up spacetime with quantum entanglement,''
  Gen.\ Rel.\ Grav.\  {\bf 42}, 2323 (2010)
  [Int.\ J.\ Mod.\ Phys.\ D {\bf 19}, 2429 (2010)]
  [arXiv:1005.3035 [hep-th]].
  %%CITATION = ARXIV:1005.3035;%%
  %39 citations counted in INSPIRE as of 27 Apr 2013



%\cite{Hartman:2013qma}
\bibitem{Hartman:2013qma}
  T.~Hartman and J.~Maldacena,
  ``Time Evolution of Entanglement Entropy from Black Hole Interiors,''
  arXiv:1303.1080 [hep-th].
  %%CITATION = ARXIV:1303.1080;%%
  %4 citations counted in INSPIRE as of 07 May 2013

%\cite{Morrison:2012iz}
\bibitem{Morrison:2012iz}
  I.~A.~Morrison and M.~M.~Roberts,
  ``Mutual information between thermo-field doubles and disconnected holographic boundaries,''
  arXiv:1211.2887 [hep-th].
  %%CITATION = ARXIV:1211.2887;%%
  %2 citations counted in INSPIRE as of 29 May 2013


%\cite{Marolf:2012xe}
\bibitem{Marolf:2012xe}
  D.~Marolf and A.~C.~Wall,
  ``Eternal Black Holes and Superselection in AdS/CFT,''
  Class.\ Quant.\ Grav.\  {\bf 30}, 025001 (2013)
  [arXiv:1210.3590 [hep-th]].
  %%CITATION = ARXIV:1210.3590;%%
  %7 citations counted in INSPIRE as of 30 May 2013

  %\cite{Marolf:2008tx}
\bibitem{Marolf:2008tx}
  D.~Marolf,
  ``Black Holes, AdS, and CFTs,''
  Gen.\ Rel.\ Grav.\  {\bf 41}, 903 (2009)
  [arXiv:0810.4886 [gr-qc]].
  %%CITATION = ARXIV:0810.4886;%%
  %13 citations counted in INSPIRE as of 30 May 2013


\bibitem{Sen:2011cn}
  A.~Sen,
  ``State Operator Correspondence and Entanglement in $AdS_2/CFT_1$,''
  Entropy {\bf 13}, 1305 (2011)
  [arXiv:1101.4254 [hep-th]].
  %%CITATION = ARXIV:1101.4254;%%
  %18 citations counted in INSPIRE as of 07 May 2013


    \bibitem{Page:1993wv}
D.~N.~Page,
 ``Average entropy of a subsystem,''
  Phys.\ Rev.\ Lett.\  {\bf 71}, 1291 (1993)
  [gr-qc/9305007].
  %%CITATION = GR-QC/9305007;%%

 D.~N.~Page,
``Information in black hole radiation,''
  Phys.\ Rev.\ Lett.\  {\bf 71}, 3743 (1993)
  [hep-th/9306083].
  %%CITATION = HEP-TH/9306083;%%


%\cite{Susskind:2013tg}
\bibitem{Susskind:2013tg}
  L.~Susskind,
 ``Black Hole Complementarity and the Harlow-Hayden Conjecture,''
  arXiv:1301.4505 [hep-th].
  %%CITATION = ARXIV:1301.4505;%%
  %3 citations counted in INSPIRE as of 12 Apr 2013

 \bibitem{Shenker:2013pqa}
  S.~H.~Shenker and D.~Stanford,
  ``Black holes and the butterfly effect,''
  arXiv:1306.0622 [hep-th].
  %%CITATION = ARXIV:1306.0622;%%

\bibitem{StanfordPrivate}
 D. Stanford, private communication. To appear. 

   %\cite{Hayden:2007cs}
\bibitem{Hayden:2007cs}
  P.~Hayden and J.~Preskill,
 ``Black holes as mirrors: Quantum information in random subsystems,''
  JHEP {\bf 0709} (2007) 120
  [arXiv:0708.4025 [hep-th]].
  %%CITATION = ARXIV:0708.4025;%%


%\cite{Verlinde:2012cy}
\bibitem{Verlinde:2012cy}
  E.~Verlinde and H.~Verlinde,
  ``Black Hole Entanglement and Quantum Error Correction,''
  arXiv:1211.6913 [hep-th].
  %%CITATION = ARXIV:1211.6913;%%
  %7 citations counted in INSPIRE as of 15 May 2013

\bibitem{HaydenPrivate}
P. Hayden, private communication.

\bibitem{PreskillBook}
J. Preskill, { \rm http://www.theory.caltech.edu/people/preskill/ph229/\#lecture }


 %\cite{Swingle:2012wq}
\bibitem{Swingle:2012wq}
  B.~Swingle,
  ``Constructing holographic spacetimes using entanglement renormalization,''
  arXiv:1209.3304 [hep-th].
  %%CITATION = ARXIV:1209.3304;%%
  %7 citations counted in INSPIRE as of 14 May 2013


%\cite{Polchinski:1999yd}
\bibitem{Polchinski:1999yd}
  J.~Polchinski, L.~Susskind and N.~Toumbas,
  ``Negative energy, superluminosity and holography,''
  Phys.\ Rev.\ D {\bf 60}, 084006 (1999)
  [hep-th/9903228].
  %%CITATION = HEP-TH/9903228;%%
  %59 citations counted in INSPIRE as of 17 May 2013






  \end{thebibliography}
\end{document}